\documentclass[11pt]{article}
\usepackage[T1]{fontenc}
\usepackage{lmodern}
\usepackage[margin=1in]{geometry}
\usepackage{amsmath,amssymb,amsthm,mathtools}
\usepackage{booktabs,array,enumitem,microtype,xcolor,float}
\usepackage[numbers,sort&compress]{natbib}
\usepackage{tikz}
\usetikzlibrary{arrows.meta,positioning,calc,fit}
\usepackage[colorlinks=true,linkcolor=blue!45!black,citecolor=blue!45!black,urlcolor=blue!45!black]{hyperref}
\hypersetup{pdftitle={Constant-Factor Matching-Size Estimation in Random-Order Streams},pdfauthor={Alireza Farhadi, Mohammad Hajiaghayi, Morteza Monemizadeh}}
\setlist{itemsep=2pt,topsep=4pt}
\newtheorem{theorem}{Theorem}[section]
\newtheorem{lemma}[theorem]{Lemma}
\newtheorem{proposition}[theorem]{Proposition}
\newtheorem{corollary}[theorem]{Corollary}
\theoremstyle{definition}

\theoremstyle{remark}

\newcommand{\E}{\mathbb E}
\newcommand{\Prb}{\mathbb P}
\newcommand{\ind}{\boldsymbol 1}
\newcommand{\KL}{D_{\mathrm{KL}}}
\newcommand{\TV}{\operatorname{TV}}
\newcommand{\ent}{\mathsf H}
\newcommand{\IID}{\mathrm{iid}}
\newcommand{\WOR}{\mathrm{WOR}}
\newcommand{\PASS}{\mathsf{PASS}}
\newcommand{\FAIL}{\mathsf{FAIL}}
\newcommand{\PAD}{\mathsf{PAD}}
\newcommand{\IDLE}{\mathsf{IDLE}}
\newcommand{\NULL}{\mathsf{NULL}}
\newcommand{\Dead}{\mathsf{DEAD}}

\newcommand{\Unif}{\operatorname{Unif}}
\newcommand{\calL}{\mathcal L}
\newcommand{\polylog}{\operatorname{polylog}}
\newcommand{\ceil}[1]{\left\lceil #1\right\rceil}
\newcommand{\floor}[1]{\left\lfloor #1\right\rfloor}

\newcommand{\proc}[1]{\textnormal{\scshape #1}}
\floatstyle{ruled}
\newfloat{algorithm}{tbp}{loa}
\floatname{algorithm}{Algorithm}
\newcommand{\dyad}{\operatorname{dyad}}
\definecolor{navy}{HTML}{234E70}
\definecolor{lightblue}{HTML}{E7EFF6}
\definecolor{lightgray}{HTML}{F0F1F3}
\title{Constant-Factor Matching-Size Estimation\\in Random-Order Streams}
\author{
Alireza Farhadi\\
Profound
\and
Mohammad Hajiaghayi\thanks{\texttt{hajiagha@umd.edu}}\\
University of Maryland
\and
Morteza Monemizadeh\thanks{\texttt{m.monemizadeh@tue.nl}}\\
TU Eindhoven
}

\date{}
\begin{document}
\maketitle
\begin{abstract}
We give a one-pass algorithm that estimates the maximum matching size of
an arbitrary simple graph within a constant factor using $O(\log^2 n)$
bits of space. The edges arrive exactly once in uniformly random order.
The algorithm succeeds with probability at least $2/3$, does not require
the number of edges in advance, and has worst-case total running time
$O((m+1)\log(n+2))$ on a word RAM.

Kapralov, Khanna, and Sudan (SODA 2014) obtained a polylogarithmic
approximation in polylogarithmic space. Kapralov, Mitrovi\'c,
Norouzi-Fard, and Tardos (SODA 2020) subsequently gave a constant-factor
approximation from independent uniform edge samples and an
$O(\log^2 n)$-approximation in random order. We obtain a constant factor
in the latter model with the same $O(\log^2 n)$ space bound.

The main ingredient is a comparison between the memory states of a
recursive sampling algorithm under sampling with and without replacement.
We account jointly for the sampled edges retained by the recursion and
the positions at which they arrived. Randomized thresholds limit the
additional information stored in numerical counters. A bound on the
expected work of the entire recursion then removes the loss from its
logarithmic depth. The resulting estimator repeatedly improves a coarse
estimate while storing only one recursion stack.
\end{abstract}

\section{Introduction}
\label{sec:intro}

How much memory is needed to estimate the size of a maximum matching
when the edges of a graph arrive in random order? A greedy algorithm
stores a maximal matching in $O(n)$ words and obtains a factor-two
estimate. If the algorithm need only output a number, however, much
less space can suffice. The question is whether a constant-factor
estimate is possible with only polylogarithmic space on arbitrary
graphs.

Kapralov, Khanna, and Sudan~\cite{KKS14} initiated this problem in the
random-order model, obtaining a polylogarithmic approximation using
polylogarithmic space. Kapralov, Mitrovi\'c, Norouzi-Fard, and
Tardos~\cite{KMNT20}, abbreviated KMNT, obtained a constant-factor
approximation using $O(\log^2 n)$ bits from at most $m$ independent
uniform edge samples. Their random-order result gives an
$O(\log^2 n)$-approximation in the same space; its stated theorem assumes
$m\ge3n$.

The distinction between the two input models matters. With independent
sampling, every observation is uniform over the entire edge set, and
an edge may appear repeatedly. In random order, each edge appears
exactly once; after a prefix has been read, the next edge is uniform
over the unread edges. A recursive algorithm therefore cannot treat
later observations as fresh samples from the original graph. For
example, after observing an edge $e$, the next sample equals $e$ with
probability $1/m$ in the independent model and with probability zero
in a random permutation. The challenge is to control these dependencies
without storing the observed prefix.

\subsection{Our result}

A matching is a set of pairwise vertex-disjoint edges. Write $\nu(G)$
for the maximum cardinality of a matching in $G$, with $\nu(G)=0$ when
the edge set is empty. For an integer $n\ge0$, let
$[n]=\{1,\ldots,n\}$, with $[0]=\varnothing$.

\begin{theorem}[Main result]\label{thm:main}
There are an absolute constant $C>0$ and a randomized one-pass
streaming algorithm such that, for every integer $n\ge0$ and every
simple undirected unweighted graph $G=([n],E)$, the following holds.
Given $n$ and a uniformly random permutation of the $m=|E|$ edges,
the algorithm outputs an integer $\widehat\nu$ satisfying
\[
 \Prb\!\left\{\frac{\nu(G)}{C}\le\widehat\nu\le\nu(G)\right\}
 \ge\frac23.
\]
The probability is over the edge order and the algorithm's independent
random bits. On every execution, the algorithm uses
$O(\log^2(n+2))$ bits of working memory and
$O((m+1)\log(n+2))$ word operations on a RAM with
$\Theta(\log(n+2))$-bit words. It does not require $m$ in advance.
It outputs zero on the empty stream and a positive integer otherwise.
\end{theorem}

The theorem applies without restrictions on degree, arboricity,
density, or matching size. It estimates the cardinality of a maximum
matching; it does not output matching edges. Table~\ref{tab:comparison}
compares the relevant estimation guarantees. The main proof gives
$C=2^{142}$. Appendix~\ref{app:sharp} sharpens the analysis to
$C=2^{28}$ without changing the asymptotic bounds. We separate that
calculation from the main argument so that the proof of a constant
approximation does not depend on optimizing constants.

\begin{table}[t]
\centering\small
\setlength{\tabcolsep}{4pt}
\renewcommand{\arraystretch}{1.2}
\begin{tabular}{@{}p{.23\linewidth}p{.34\linewidth}p{.18\linewidth}p{.17\linewidth}@{}}
\toprule
Result & Input model & Approximation & Space (bits)\\
\midrule
KKS~\cite{KKS14} & Random-order stream & $\polylog n$ & $\polylog n$\\
KMNT~\cite[Theorem 1]{KMNT20} & At most $m$ independent edge samples
  & $O(1)$ & $O(\log^2 n)$\\
KMNT~\cite[Theorem 4]{KMNT20} & Random-order stream, $m\ge3n$
  & $O(\log^2 n)$ & $O(\log^2 n)$\\
\midrule
Theorem~\ref{thm:main} & Random-order stream & $O(1)$ & $O(\log^2 n)$\\
\bottomrule
\end{tabular}
\caption{One-pass estimation of maximum matching size. The KMNT theorem
numbers refer to the full version. All graphs are simple.}
\label{tab:comparison}
\end{table}

\paragraph{Computational model.}
The algorithm receives $n$ and then reads the edges, followed by an
end-of-stream marker. Its word length is fixed from $n$ alone and is
large enough for vertex labels and the signed integer quantities used
below. It uses standard integer arithmetic, comparisons, addressable
working memory, and fresh fair random bits independent of the stream.
The time bound counts stream reads and random-bit draws; the space
bound includes registers and active memory words. Both bounds, as
well as termination and the stated output-sign guarantees, hold
uniformly for every graph, edge permutation, and coin sequence.

\paragraph{Main ideas.}
Our algorithm builds on recursive peeling~\cite{KKS14,KMNT20}.
A recursive test explores whether a vertex remains after a specified
number of peeling levels. An edge initiates two such tests, one for
each endpoint. Their outcome defines a numerical reward whose mean
approximates the matching size. The implementation stores the active
tests in a single stack.

The principal difficulty is to compare this computation under
independent samples and under a random permutation. A comparison of
the entire observed sequence detects repeated edges and is too costly.
We instead bound the information retained in the current memory
state, using relative entropy.\footnote{For discrete probability laws
$P,Q$, their relative entropy is
$D_{\mathrm{KL}}(P\Vert Q)=\sum_xP(x)\log(P(x)/Q(x))$, with the
usual zero-mass conventions and value $+\infty$ if $Q(x)=0<P(x)$.
All information quantities in this paper use natural logarithms.}
The state retains a small number of sampled edges, their arrival
positions, and randomized residual counters. Accounting for the
positions and labels jointly avoids a separate $\log n$ charge for
each stored identity. An occupation bound for the recursion shows
that the expected total information charge of one edge experiment is
proportional to its expected reward, independently of the number of
levels. Section~\ref{sec:techniques} explains these statements and how
they yield the estimator.

\subsection{Related work}

\paragraph{Sparse graphs and local algorithms.}
Esfandiari, Hajiaghayi, Liaghat, Monemizadeh, and
Onak~\cite{EHLMMO15} gave constant-factor matching-size estimators
in sublinear space for planar and bounded-arboricity graphs in
arbitrary order. Cormode, Jowhari, Monemizadeh, and
Muthukrishnan~\cite{CJMM17} obtained polylogarithmic-space estimators
for sparse graphs. McGregor and Vorotnikova~\cite{MV18} gave an
$(a+2)(1+\varepsilon)$-approximation for arboricity-$a$ graphs using
$O(\varepsilon^{-2}\log n)$ words in arbitrary insertion order.
For bounded-degree graphs, Monemizadeh, Muthukrishnan, Peng, and
Sohler~\cite{MMPS17} developed random-order simulations of local
property testers, with consequences for matching estimation that
allow additive error. Our theorem requires none of these structural
promises.

\paragraph{Weighted and dynamic estimation.}
Bury and Schwiegelshohn~\cite{BS15} reduced estimation of maximum
matching weight to estimation of unweighted matching size. Their
reduction applies to dynamic streams, where edges may be deleted,
and they obtained a constant-factor estimator for bounded-arboricity
graphs in that model using $\widetilde O(n^{4/5})$ space. In a
different setting, dynamic algorithms maintain a matching-size
estimate after each update; see, for example,
Behnezhad~\cite{DBLP:conf/soda/Behnezhad23}.

\paragraph{Constructing a matching.}
A separate line of work returns an actual matching in one random-order
pass using near-linear space. Bernstein~\cite{Bernstein20} obtained
a $(2/3-\varepsilon)$-approximation, and Assadi and
Behnezhad~\cite{AB21} improved the ratio to $2/3+\varepsilon_0$ for
an absolute $\varepsilon_0>0$. A recent preprint of Farhadi,
Hajiaghayi, and Mahdavi~\cite{FHM26} reports a $0.71295$-approximation
for general unweighted graphs in near-linear space. These results
optimize the size of the matching returned. Our goal is a
constant-factor estimate using polylogarithmic memory.

\paragraph{Lower bounds and information-theoretic methods.}
Assadi, Khanna, and Li~\cite{AKL17} proved space--approximation
tradeoffs for matching-size estimation in insertion-only and dynamic
streams. Assadi and Sundaresan~\cite{AS23} proved random-order lower
bounds in fine-additive and near-exact multiplicative regimes.
Efremenko, Kol, Saxena, and Zhang~\cite{EKSZ26} established lower
bounds against white-box adaptive adversaries. These results concern
different accuracy requirements or input models; they do not establish
an $\Omega(\log^2 n)$ lower bound for the problem studied here.
Our state-distribution comparison is related to streaming switching
arguments, including Shahaf, Ordentlich, and Segev~\cite{SOS20}.
The bound used here depends on the structure and expected occupation
of the retained state, rather than only on its worst-case bit length.

\section{Our approach and techniques}
\label{sec:techniques}

We describe the proof in five steps. First, recursive tests turn a
uniformly sampled edge into a reward related to $\nu(G)$. Second,
fixed return times allow those tests to run on one stream. Third,
we bound the total work of the recursion, including the logarithmic
weights needed for the probability comparison. Fourth, we compare
the memory state under the two input distributions. Finally, a
completion count and a sequence of refinement stages yield the
estimate. Throughout this overview, $m>0$ is fixed and all
unspecified constants are absolute.

\subsection{Recursive tests and the matching estimate}

The starting point is the peeling method of KMNT~\cite{KMNT20}.
At progressively larger sampling scales, the algorithm tests whether
a vertex has few neighbors that survive earlier tests. Vertices in
dense regions tend to fail earlier; a vertex with few surviving
neighbors can pass additional levels. We use this principle with
different stopping and timing rules to support the random-order
analysis.

A \emph{call} is one invocation of the procedure
$\proc{Vertex}(v,j)$, which returns $\PASS$ or $\FAIL$ for vertex
$v$ at level $j\ge1$. Its sampling scale is $K_j=c^{j-1}$, where
$c>1$ is a fixed power of two. On an incident edge $\{v,w\}$,
a level-$j$ call tests $w$ at levels $1,\ldots,j-1$, stopping at the
first failure. These are its \emph{child calls}; the call on $v$ is
their \emph{parent}. It subtracts the resulting contribution from
a random residual counter and fails when the counter reaches zero.
The exact rules appear in Algorithm~\ref{alg:vertex}.

For the moment, analyze these calls on independent uniform edge
samples. Let $q_j(v)$ be the probability that a full level-$j$ call
passes, and let
\[
 p_0(v)=1,\qquad p_i(v)=\prod_{j=1}^i q_j(v).
\]
Thus $p_i(v)$ is the probability that fresh tests of $v$ pass all
levels through $i$. A level is an index of a randomized test, not
a vertex set stored by the algorithm. The probabilities $p_i(v)$
are used only in the analysis.

A \emph{root} is an experiment started by a uniform edge $uv$.
It tests $u$ and $v$ at level one, then at level two, and so on,
until a test fails or both endpoints pass the maximum level $J$.
If $I$ is the last level passed by both endpoints, the reward is
$H_I=1+c+\cdots+c^I$. In particular, $I=0$ and the reward is one
if a level-one test fails. The \emph{cap} $h=H_J$ is the largest
possible reward. Since $H_I=\sum_{i=0}^J c^i\ind_{\{I\ge i\}}$,
\[
 \mu:=\E[H_I]
 =\frac1m\sum_{uv\in E}\sum_{i=0}^J c^ip_i(u)p_i(v).
\]
Proposition~\ref{iid:proxy} proves
\[
 \mu=O(\nu(G)),\qquad
 \mu=\Omega\!\left(\frac hm\nu(G)\right)\quad\text{when }h\le m.
\]
These inequalities are the link between sampling and matching. An
estimate of $\mu$ never overestimates $\nu(G)$ by more than a
constant, and a larger cap improves the approximation from below.

Here is the matching argument. Give each edge its summand in the
formula for $\mu$ as a weight. Once a vertex has accumulated a large
load, its probability of passing subsequent levels decreases rapidly;
this bounds its total weighted degree by a constant. The endpoints
of any maximum matching form a vertex cover, giving the upper bound
on the sum of edge weights. For the lower bound, consider vertices
with $p_J(v)\le1/2$. Each has accumulated a constant weighted degree
before reaching that probability. If there are many such vertices,
their contributions suffice. Otherwise a constant fraction of the
edges in a fixed maximum matching have both endpoints with survival
probability above $1/2$, and their terms through level $J$ contribute
$\Omega(h/m)$ each.

The survival estimate used here does not lose a factor of $J$.
At a given scale, split a comparison score into its integer
contribution at the highest level and the contribution inherited
from lower levels. The first is nonzero with probability at most
its mean. The second crosses the threshold only through a deviation
controlled by its variance. A load introduced at level $a$ contributes
to this variance at level $i>a$ with weight $c^{a-i}$. Summing over
later levels costs at most $1/(c-1)$, so the sum of failure
probabilities is bounded by the total load itself.

\subsection{Fixed return times for recursive calls}

A recursive call uses part of its parent's stream interval. Returning
only the children that happen to finish early could bias the parent
toward particular outcomes. We avoid this problem by fixing a call's
return time before it examines its samples.

A level-$j$ call invoked at boundary $\sigma$ waits until the next
multiple $a$ of $K_j$, examines its RUN interval of $K_j$ positions,
and returns at $a+K_j$. We call the interval before $a$ WAIT. Even
if the answer is decided early, the call discards subsequent edges
and returns only at its assigned deadline. Its total duration is at
most $2K_j$.

Every higher-level scale is divisible by $K_j$. An ancestor's
deadline is therefore a multiple of the child's scale. If that
deadline precedes the child's return time, it is at or before the
start of the child's RUN interval. The child is then canceled before
reading any RUN edge. Otherwise it executes its full test. Under
independent sampling, a completed call consequently has a law that
depends only on $(v,j)$, regardless of its invocation time.

\begin{figure}[t]
\centering
\begin{tikzpicture}[x=.82cm,y=.7cm,font=\small,>=Stealth]
\draw[->] (0,0)--(10.5,0) node[right]{time};
\foreach \x in {0,2,4,6,8,10}
  {\draw (\x,.1)--(\x,-.1) node[below=3pt]{$\x$};}
\node[anchor=east] at (-.2,3.3){parent};
\filldraw[fill=lightblue,draw=navy] (0,2.9) rectangle (8,3.7);
\node at (4,3.3){RUN, deadline $8$};
\draw[dashed,gray] (8,.25)--(8,4);
\node[anchor=east] at (-.2,2){child};
\filldraw[fill=lightgray,draw=gray] (1,1.6) rectangle (2,2.4);
\filldraw[fill=lightblue,draw=navy] (2,1.6) rectangle (4,2.4);
\node[font=\scriptsize] at (1.5,2){WAIT};
\node at (3,2){RUN};
\node[anchor=west,font=\scriptsize] at (4.2,2){returns its full test at $4$};
\node[anchor=east] at (-.2,.75){late child};
\filldraw[fill=lightgray,draw=gray] (7,.35) rectangle (8,1.15);
\node[font=\scriptsize] at (7.5,.75){WAIT};
\draw[dashed,navy] (8,.35) rectangle (10,1.15);
\node[font=\scriptsize,align=center] at (9,.75){canceled};
\node[anchor=east,font=\scriptsize] at (6.8,.75){invoked at $7$};
\end{tikzpicture}
\caption{Fixed deadlines prevent completion from selecting favorable
outcomes. This schematic uses parent scale $8$ and child scale $2$;
the algorithm uses a larger ratio of consecutive scales.}
\label{fig:calendar}
\end{figure}
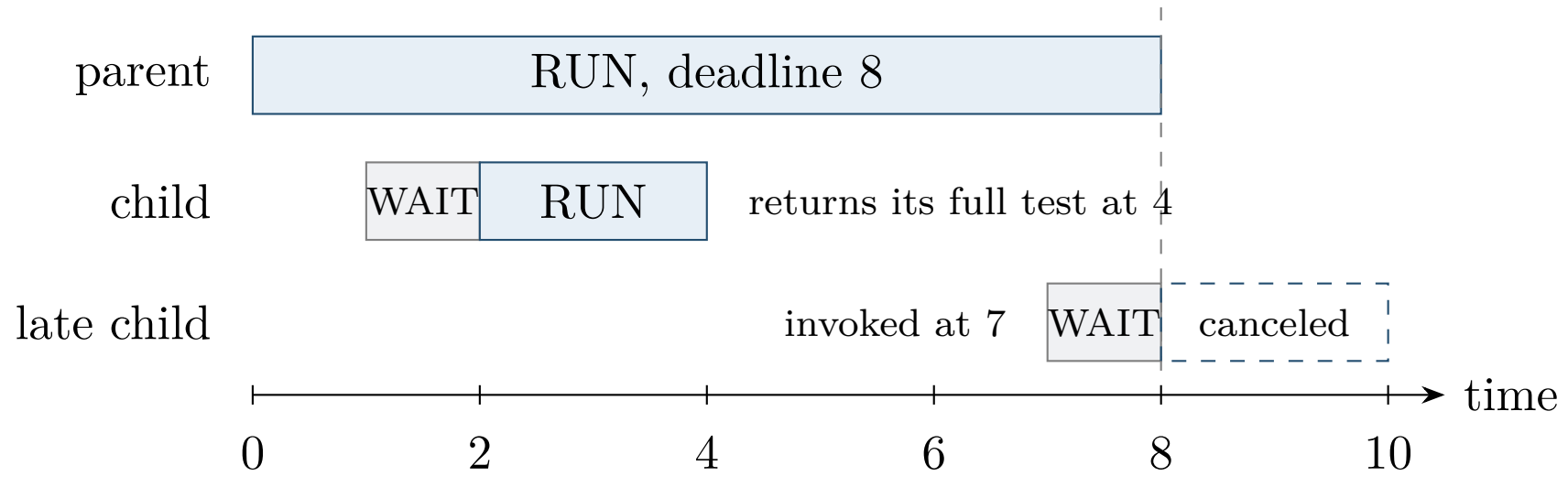

At each position, only the deepest active call may examine the edge;
this owner is chosen before the edge is read. A waiting parent does
not reuse its child's edge. At a common deadline, completed child
answers and immediate failure rules are processed before a parent's
default answer. These conventions make the recursive tests a
one-pass streaming procedure.

\subsection{Bounding the work of the entire recursion}

The matching estimate alone does not control the information stored
by the stack. Retaining a vertex that rarely survives can reveal
more information than retaining a common outcome. We must bound
the work of the recursion with weights such as $\log(1/p_i(v))$.

The sampling budget of a call is $K_j$; its duration, which includes
waiting, is at most $2K_j$. If a neighbor's first failed test is
at level $i$, its child calls have combined budget
$K_1+\cdots+K_i=H_{i-1}$, exactly the contribution subtracted from
the parent's residual. The stopping rule bounds the sum of all
direct-child budgets by $3\delta K_j$ on every execution, for a
small constant $\delta$.

Index call types by $(i,v)$. Let $T_{xy}$ be the expected budget
of children of type $x$, per unit of budget at a parent of type
$y$. The preceding bound gives column sums at most $3\delta$.
There is also an initial budget vector $\omega$, obtained from the
root's endpoint tests, such that $T\omega\le2\omega$ and
$\sum_x\omega_x=O(\mu)$. This second bound uses the fact that a
neighbor passing all required lower levels makes the parent fail.
The corresponding failure probabilities telescope across levels.

The two inequalities imply that $T$ contracts the weighted Euclidean
norm $\|f\|_{L^2(\omega)}^2=\sum_x\omega_x f_x^2$ after the
appropriate change of coordinates. Consequently the expected budget
over all depths, bounded by
$\omega+T\omega+T^2\omega+\cdots$, has a bounded norm independent
of the depth. Cauchy--Schwarz and the inequality
$p\log^2(1/p)\le3\sqrt p$ then control the logarithmic weights.
Proposition~\ref{iid:fee} concludes that the expected total
information charge of a root is $O(\mu)$. Here a charge is a
nonnegative quantity assigned in the analysis to a call or root
entry; it bounds that record's contribution to the later state
comparison. The algorithm does not compute these charges.

\subsection{Comparing the retained memory states}

Let $A_t$ be the set of the first $t$ edges of a uniform permutation,
and let $S_t$ be the memory state before the next edge is read.
The next edge is uniform on $E\setminus A_t$. A uniform-subset
entropy inequality shows that choosing one element from a set of
size $m-t$ reduces the relevant relative entropy by a factor
$1/(m-t)$.\footnote{Precisely, if $\xi$ is a law on $b$-element
subsets of an $M$-element set and $K$ chooses a uniform member,
then $D_{\mathrm{KL}}(\xi K\Vert\Unif[M])\le
b^{-1}D_{\mathrm{KL}}(\xi\Vert\Unif\binom{[M]}b)$.
Equation~\eqref{tr:eq:slice} gives the proof.}
We therefore seek a bound on how much the retained state reveals
about the prefix set.

The identities in the active stack come from the root edge and the
edges that initiated pending neighbor tests. Call these edges,
together with their arrival positions, the \emph{source anchors}.
For a fixed observation time, we construct a reference distribution
by first proposing an anchor chain, then sampling a uniform
permutation subject to those edge labels at those positions. The
proposal weights use the survival probabilities of the tests needed
to reach the chain. When $r$ distinct edge labels are prescribed,
the likelihood contains the falling factorial $(m)_r$. This factor
cancels the uniform-edge factors in the proposal. The remaining
terms are survival logarithms and normalizing constants, which are
bounded by the charges from the preceding subsection.

Numerical counters require another argument. A call starts with a
uniform threshold $U\in\{M,\ldots,2M-1\}$ and stores only the
residual $U-s$, or a decided flag after crossing zero. For a fixed
integer score $s\ge0$ independent of $U$, the residual law is
dominated by three times a reference law that does not depend on
$s$. Thus its information cost is constant, even when $M$ is large.
To use this observation without assuming independence after
conditioning on the actual stack, we reconstruct each current
score with that call's threshold test suppressed. An ancestor's
reconstruction stops just before the edge that started its pending
child. These reconstructions avoid all current threshold coordinates,
and agree with the actual scores whenever the proposed chain is
the actual chain. Section~\ref{sec:transfer} proves this for every
descriptor in the reference distribution.

\begin{figure}[t]
\centering
\begin{tikzpicture}[font=\small,>=Stealth,
 box/.style={draw=navy,align=center,inner sep=7pt,text width=4.7cm},
 arr/.style={->,draw=navy,thick}]
\node[box,fill=lightblue] (actual) at (0,2.7)
  {Actual memory state\\anchors, residuals, flags};
\node[box,fill=lightgray] (proposal) at (6.7,2.7)
  {Reference distribution\\propose anchors and positions};
\node[box] (ghost) at (0,.6)
  {Reconstruct scores\\without current thresholds};
\node[box] (slice) at (6.7,.6)
  {Replace residual laws\\prefix uniform given\\anchor labels};
\draw[arr] (actual.south)--(ghost.north);
\draw[arr] (proposal.south)--(slice.north);
\draw[arr] (ghost.east)--node[above,font=\scriptsize,align=center]
  {bounded\\density ratio}(slice.west);
\node[draw=navy,align=center,inner sep=7pt,text width=10.7cm]
  (bound) at (3.35,-1.45)
  {Uniform-subset entropy inequality\\bounds dependence of the next edge on the state};
\draw[arr] (slice.south)--(6.7,-.55)--(bound.north -| slice.south);
\end{tikzpicture}
\caption{The reference distribution is used to bound information at one
time. It is constructed separately at each time and is never sampled by
the streaming algorithm.}
\label{fig:reference}
\end{figure}
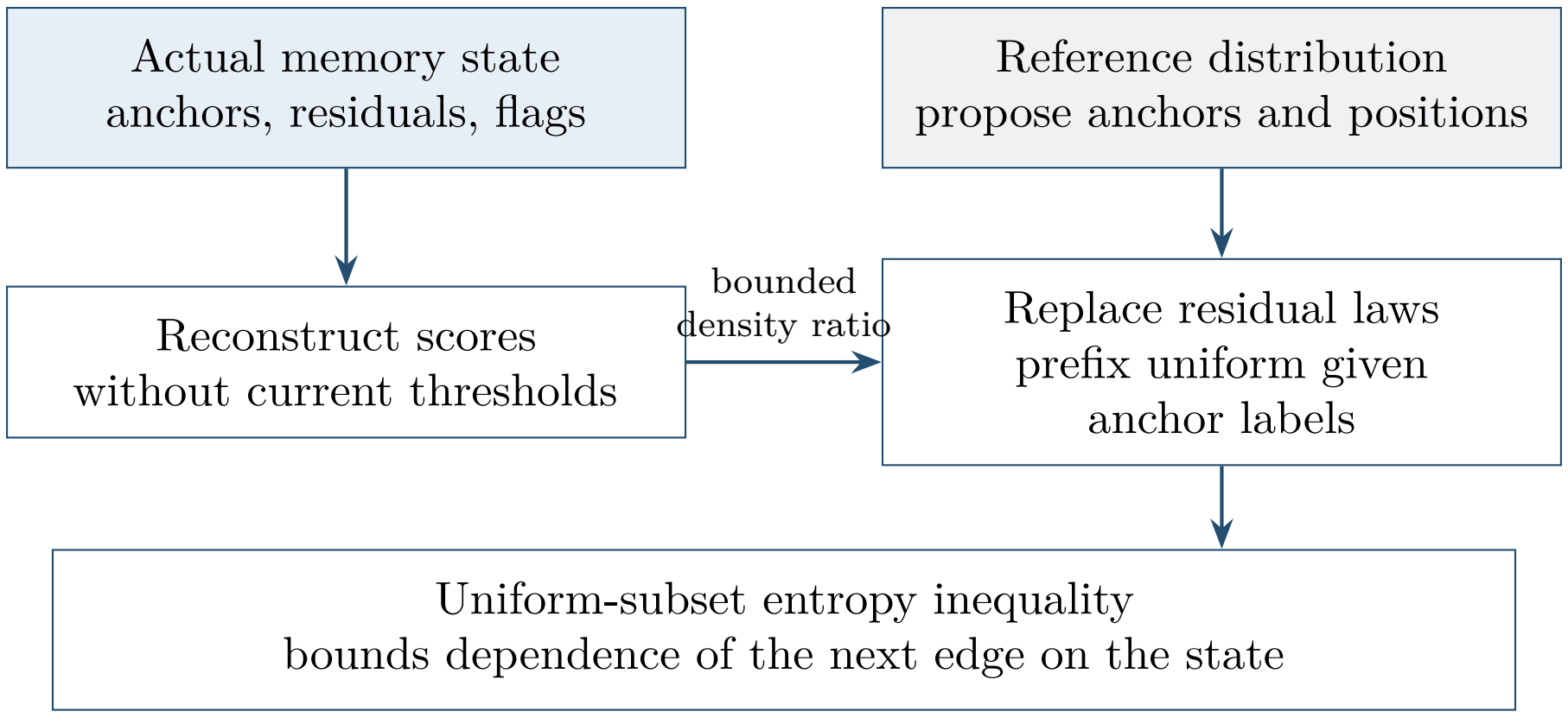

After replacing these residual laws, the reference prefix is uniform
subject only to the anchor labels. Theorem~\ref{tr:thm:anchors}
converts this fact into a bound on the information between $S_t$
and the next edge. Applying the common state-transition rule and
summing over times bounds the relative entropy between the terminal
states under independent samples and random order.

\subsection{Completion counts and successive refinement}

We now turn the reward estimate into a streaming output. A root with
reward $H$ is padded to last exactly $7H$ positions, including its
birth edge. When no root is active, the algorithm starts one with
probability $1/L$ before seeing the next edge. The expected duration
of an idle wait and a root is therefore $\Theta(L+\mu)$. During
a \emph{phase} of $T$ consecutive positions, the number $N_T$ of
completed roots is, with constant probability,
$\Theta(T/(L+\mu))$. Taking $h$ to be a small constant times $T$
gives the concentration needed for this statement.

The exact counter $N_T$ is read only at the end of a phase and
never changes the sampler's decisions. We can therefore omit it
from the state comparison. For analysis, an independent tester
counts a randomly thinned subset of completions in a counter with
constant capacity. Its acceptance certifies the same constant-factor
interval for $N_T$ under either input law. An additional monitor,
also used only in the proof, terminates an execution at a small
charge-dependent rate. The expected-charge bound makes such
termination unlikely under independent sampling, while the monitor
bounds the accepted charge under either input distribution.
Transferring the joint event of survival and tester acceptance
establishes the completion estimate in random order.

Suppose the current estimate satisfies $L\le\nu(G)\le FL$.
A phase using $T=\Theta(m/(1+\log F))$ positions has
$h=\Theta(T)$ and produces a new lower estimate within
$O(1+\log F)$ of $\nu(G)$. We apply this improvement to a
predetermined decreasing sequence of approximation targets, using
consecutive disjoint blocks of the stream. The stage lengths and
repetition counts are chosen so that all stages together read only
a small constant fraction of the edges and the sum of their failure
probabilities is less than a fixed constant. The stages run
sequentially and reuse the same recursion stack.

A short random prefix removes only a small fraction of a fixed
maximum matching with high probability. At the start of each phase,
we condition on its complete past and analyze the uniformly ordered
unread graph. We use the matching-preservation event only when
combining these conditional guarantees, avoiding conditioning on a
future success event.

Finally, the known-length algorithm fixes its answer after a short
prefix on every execution. For unknown length, we run instances for
prescribed geometrically increasing sizes, with disjoint short
active intervals, and return the most recent answer committed before
the stream ends. The last committed size depends only on the stream
length. Its observed block can be completed in the analysis to a
uniform edge-subgraph of almost the original size. This gives the
unknown-length result with one active instance and only logarithmic
additional space.

\section{The recursive streaming sampler}
\label{sec:algorithm}

A phase runs the recursive sampler for a fixed number $T$ of consecutive
stream positions and estimates a quantity from the number of completed
roots. We first specify that sampler. Sections~\ref{sec:refinement}
and~\ref{sec:unknown} choose the phases and handle unknown stream length.
Every phase starts with a fresh local clock at boundary zero.
All logarithms in probability and information bounds are natural;
$\log_2$ is used for integer schedules.

\paragraph{Scales and thresholds.}
Fix the dyadic constants $c>1$, $\delta>0$, and $0<\theta<1$,
with $c$ an integer power of two. In the main proof we use
$(c,\delta,\theta)=(2^{12},2^{-5},2^{-165})$; all other constants
are collected in Section~\ref{sec:parameters}. Put
\[
 K_j=c^{j-1}\quad(j\ge1),\qquad
 H_i=\sum_{a=0}^ic^a\quad(i\ge0),\qquad
 M_j=\max\{1,\delta K_j\}.
\]
Here $K_j$ is the length of a level-$j$ RUN interval, $H_i$ is the
reward after both root endpoints pass level $i$, and $M_j$ is the
scale of the initial residual counter. For the stated parameters,
every $M_j$ is an integer power of two. In particular,
\begin{equation}
 H_{j-2}<K_j/(c-1)<\delta K_j\le H_{j-1}
 \quad(j\ge2).\label{alg:guard}
\end{equation}
This inequality identifies when a neighbor classification reaches the
threshold scale: its geometric prefix remains below $M_j$ until
all $j-1$ lower levels have passed.

\subsection{Calls, classifications, and ownership of an edge}

A call is an invocation of the vertex test below. A neighbor
classification is a sequence of such calls at increasing levels.
At an integer physical boundary $\sigma\ge0$, a call $\proc{Vertex}(v,j)$ sets
$a=K_j\lceil\sigma/K_j\rceil$ and $d=a+K_j$. It waits until $a$,
runs on positions $a+1,\ldots,d$, and returns at $d$.
With probability $\theta$ it decides $\PASS$ immediately and discards
edges until $d$. This fixed blind branch ensures positive survival
probabilities in the proof. An ordinary call initializes its residual
uniformly in $\{M_j,\ldots,2M_j-1\}$ and follows
Algorithm~\ref{alg:vertex}. The residual is the \emph{only} retained
score variable; its original value is not kept.
A neighbor classification tests levels $1,2,\ldots,j-1$ in order,
stopping at the first failed call. If that first failure is at level
$i$, its contribution to the parent is $H_{i-1}$. If all $j-1$ tests
pass, the parent fails immediately. Equation~\eqref{alg:guard} shows
that this is equivalent to stopping when the accumulated geometric
prefix first reaches $M_j$. At level one, any incident edge fails
the ordinary call immediately.
\begin{algorithm}[t]
\caption{$\proc{Vertex}(v,j)$ invoked at boundary $\sigma$ }
\label{alg:vertex}
\small
\begin{enumerate}[leftmargin=1.7em,label=\arabic*.,itemsep=3pt]
\item Set $a\gets K_j\lceil\sigma/K_j\rceil$, $d\gets a+K_j$.
With probability $\theta$, decide $\PASS$ and go to step 6.
\item Set $R\gets\Unif\{\lfloor M_j\rfloor,\ldots,2\lfloor M_j\rfloor-1\}$; wait to boundary $a$.
\item While undecided, before $d$, and with no child active, consume
the next physical edge. If it is not incident to $v$, continue.
Otherwise write it as $\{v,w\}$.
\item If $j=1$, decide $\FAIL$. Otherwise classify $w$ by calls
$\proc{Vertex}(w,1),\ldots,\proc{Vertex}(w,j-1)$, in order:
\begin{enumerate}[leftmargin=1.5em,label=(\alph*),itemsep=2pt]
\item On the first $\FAIL$ at level $i$, set $R\gets R-H_{i-1}$,
erase the neighbor record, and decide $\FAIL$ if $R\le0$.
If $R>0$, return to step 3.
\item If all $j-1$ calls return $\PASS$, decide $\FAIL$.
\item If this call's deadline arrives first, discard the unfinished
classification, without adding a contribution.
\end{enumerate}
\item At boundary $d$, return $\PASS$ if still undecided.
\item After any early decision, erase $R$ and obsolete records;
discard edges until $d$, then return the decided answer.
\end{enumerate}
\vspace{2pt}
\noindent An earlier ancestor deadline cancels the call without an answer.
Immediate guards and completed child returns precede endpoint defaults.
No child or successor call is launched at an expired deadline
or at the external phase horizon.

\end{algorithm}
For precision, edge $b$ occupies $[b-1,b)$. Its owner is fixed before
it is inspected. When a root is active, only its deepest live call
can own the edge, and only during that call's ordinary, undecided RUN.
WAIT intervals and post-decision padding discard it. A parent waiting
for a child never examines the child's edge.
After reading edge $b$, advance to boundary $b$. Apply its immediate
guard, then process completed children and parent arithmetic from
deepest to shallowest. A returned PASS may advance a classification
or trigger its all-levels-PASS guard; a returned FAIL subtracts its
contribution. These operations precede the corresponding parent's
endpoint default. If an ancestor deadline is reached at boundary $b$
before a child's own deadline, cancel that child without returning a bit.
Launch a successor
only when $b$ is strictly before every remaining ancestor deadline and
the external phase horizon. A resumed parent can first read edge $b+1$.
These conventions include two boundary cases. An incident edge at a
level-one call's deadline fails that call before its default PASS.
At a higher-level deadline, the last own-source edge may create an
unfinished neighbor record, but no child is launched there; that record
is discarded. Between reads, temporary arithmetic, popped frames,
finished neighbor identities, and decision causes are erased.
\subsection{Padded roots}
Padding makes the number of completed roots reflect their mean reward.
It also erases the time and cause of an early decision from the retained
state. The padding rule below gives an exact duration for every full root.

A root starts from an edge $e=\{u,v\}$ read at position $b$.
Order $u,v$ by their vertex labels. Test $u$ and then $v$ at level
one, then at level two, and so on through the supplied cap $J$.
Stop at the first failed test or after both endpoints pass level $J$.
Let $I$ be the last level passed by both endpoints, with $I=0$ if
level one fails, and assign reward $H_I$. At cap $J=0$ there are
no calls and the reward is one.
If the root naturally finishes at boundary $d$, then
\[
 d-(b-1)\le1+4H_I\le5H_I.
\]
Indeed, a first failure at level $I+1$ requires at most two calls
per level through $I+1$, each with duration at most $2K_j$.
Reaching the cap uses no more time. We therefore pad the root to
the strictly later boundary
\begin{equation}
 D^{\rm pad}=b-1+7H_I.\label{alg:pad}
\end{equation}
During $\PAD$, discard every source edge. At $D^{\rm pad}$ count
one completion, erase the root, and become idle. Thus each full root
has duration exactly $7H_I$, including its birth edge.
The PAD state retains only $(e,b,I)$ and its mode. In particular,
the natural decision time and failed endpoint are erased.
The parameters $\kappa$ and $S$ are fixed positive constants. A phase
has an integer horizon $T$ and a dyadic integer rate $L\ge1$. All executed
phases below have $\kappa T\ge H_1$, so step 1 is well defined.
Dyadic rates make the birth coins exact. Dyadic rounding  is implemented
by integer comparisons, not floating-point logarithms. A root whose
natural decision occurs at $T$ is incomplete, since its padded
deadline is later. The proof compares the state immediately after the
cleanup in step 4; only then is the exact count used to form an estimate.
\begin{algorithm}[t]
\caption{$\proc{Phase}(L,T)$: a fixed-horizon completion estimate}
\label{alg:phase}
\small
\begin{enumerate}[leftmargin=1.7em,label=\arabic*.,itemsep=3pt]
\item Choose the largest $J\ge0$ with $H_J\le\kappa T$;
set the completion counter $N\gets0$ and start idle.
\item Process exactly $T$ physical edge positions.
When idle, toss an independent Bernoulli-$1/L$ coin \emph{before}
reading the next edge. On success, start a root on that edge;
on failure, discard it.
\item When active, execute the current root with cap $J$ and the
dispatcher above. Whenever its padded deadline is reached, increment
$N$ and become idle. The counter never affects the sampler.
\item At boundary $T$, finish all returns due there and count a
root padded exactly to $T$. Launch no new call; discard any
unfinished root and clear the stack.
\item Return $1$ if $N=0$. Otherwise return
$\max\{1,\dyad(T/(SN))\}$, where
$\dyad(x)=2^{\lfloor\log_2x\rfloor}$.

\end{enumerate}
\end{algorithm}
\paragraph{Stored state.}

The root keeps its two endpoint labels, birth position, current level,
and endpoint role. Each call frame keeps its vertex, level, calendar,
residual or a cleared value, and a constant number of flags. A pending
classification keeps its source birth and current child level; its
neighbor label is the child's vertex. Levels strictly decrease down
the stack. Thus there are $O(\log(n+2))$ frames of
$O(\log(n+2))$ bits each. The exact completion count is stored
separately and is not consulted within a phase. No pass probabilities,
loads, reference distributions, information charges, monitors, or testers
from the proof are evaluated by the algorithm.
\section{The i.i.d. matching proxy and its information cost}
\label{sec:iid}

This section establishes two bounds for the sampler under independent
uniform edge samples: the mean root reward approximates the matching
size, and the expected total information charge of a root is
proportional to that reward. The first will justify the estimator;
the second will allow its analysis to be transferred to random order.

Throughout this section, $G=(V,E)$ is a finite simple graph with
linearly ordered vertices, $m=|E|>0$, and $\nu=\nu(G)$. Source
edges are independent and uniform on $E$, and all private call coins
are independent of the source and one another. Unless a lemma states
otherwise, use the recursive-test constants from
Section~\ref{sec:parameters}. The blind-PASS probability $\theta>0$
keeps all survival probabilities positive. Recall that
$K_j=c^{j-1}$ and $H_i=\sum_{a=0}^i c^a$.

\subsection{Canonical calls and the sample sandwich}

The first lemma makes each call type well defined probabilistically.
Alignment ensures that a deadline cannot select a favorable proper
prefix of a child's RUN interval. A \emph{canonical call} is a full
call on an unlimited independent edge source with no ancestor cutoff.

\begin{lemma}[Canonical calls]\label{iid:canonical}
Let $v\in V$ and $j\ge1$ be an integer. A full level-$j$ call
invoked at any integer boundary $\sigma\ge0$ has a PASS probability
$q_j(v)\in[\theta,1]$ independent of $\sigma$.
If it is invoked beneath aligned higher-level ancestors with deadlines
after $\sigma$, an ancestor deadline strictly earlier than its scheduled return
cancels it before RUN, and it launches no children. Conditional on a positive-probability
invocation history with fresh future samples and coins, the multiset
of its direct child types, even under a further external cutoff,
is stochastically dominated by that of a full canonical call of
the same type.
\end{lemma}
\begin{proof}
The RUN start of a level-$j$ call is a multiple of $K_j$.  Translating
its RUN to any other such start preserves all lower-level calendars,
since $K_i$ divides $K_j$ for $i<j$.  Induction on $j$, independence of
future i.i.d. edges, and freshness of private coins prove the first claim.
Every ancestor deadline is a multiple of $K_j$.  If such a deadline
precedes the call's return, it is therefore at or before its RUN
start.  A canceled call produces no children.  Otherwise the call
executes its entire canonical RUN, including its boundary-discard rule.
A fixed external horizon is handled by completing the one unfinished
root on a hypothetical i.i.d. continuation.  All charges considered below
are nonnegative, so this only increases their total.
\end{proof}
For integers $i\ge0$, define the survival probabilities and loads by
\[
\begin{gathered}
 p_0(v)=1,\qquad p_i(v)=\prod_{j=1}^i q_j(v),\qquad
 D_i(v)=\sum_{w\sim v}p_i(w),\\
 a_i(v)=\frac1m\sum_{a=0}^ic^aD_a(v),\qquad a_{-1}(v)=0.
\end{gathered}
\]
Here $D_i(v)$ is the sum of the survival probabilities of the
neighbors of $v$, and $a_i(v)$ is their cumulative weighted load.
Thus $p_i(v)>0$, $a_0(v)\le1$, and
$a_i(v)\le(c+1)a_{i-1}(v)$ for $i\ge1$.  Successive calls use fresh, disjoint
source intervals and have the phase-independent laws of
Lemma~\ref{iid:canonical}.   A fresh classification of $v$ therefore
passes its first $i$ levels with probability $p_i(v)$.
For a fixed level-$j$ parent, construct a comparison tuple $(e,L)$
for each edge it examines on its own behalf. Here $e$ is uniform on
$E$ and $L=(L_w)_{w\in V}$ is independent of $e$, with independent
coordinates $L_w\in\{0,\ldots,j-1\}$ satisfying
$\Prb(L_w\ge a)=p_a(w)$ for $0\le a\le j-1$.
These tuples can be taken independent of one another and of the
parent's residual and blind-branch coin.

To justify this coupling, expose a tuple only after the parent
requests its next own-source edge. Completed classifications use
their actual test outcomes; uninspected vertices receive independent
truncated levels. Complete the last unfinished classification on an
independent continuation, and append fresh tuples after the parent
stops. There is no later own-source edge of this parent with which
that final completion can conflict. This constructs a comparison
sequence for one parent; it does not assert independence of the
continuations used for different parents.

For an integer $i\ge0$, use tuples truncated at level $i$ to define
\[
 Y_i(v,e,L)=\ind_{\{v\in e\}}
       \sum_{a=0}^ic^{a-i}\ind_{\{L_{e\setminus\{v\}}\ge a\}},
 \qquad S_i(v)=\sum_{b=1}^{c^i}Y_i(v,e_b,L_b).
\]
When $v\in e$, the notation $e\setminus\{v\}$ means the other
endpoint; otherwise the mark is zero. The tuples $(e_b,L_b)$ are
independent copies, so $0\le Y_i<2$ and $\E[S_i(v)]=a_i(v)$.
For example, if a neighbor fails first at level $a+1$, its mark
contains exactly the contributions from levels $0,\ldots,a$.

The next lemma relates this comparison sequence to the actual
execution. A failure requires a large score, whereas a passing
call must have completed many comparison samples. The same
deterministic budget bound proves both assertions.

\begin{lemma}[Work and samples]\label{iid:sandwich}
For every $j\ge1$ and every execution of a level-$j$ call on any
edge sequence, including ancestor or external-horizon truncation,
its direct children have
combined budgets below $3\delta K_j$.  This remains true if the final
unfinished neighbor classification is completed up to the prefix guard.
For $j\ge2$, a full call on $v$, conditional on its ordinary
(non-blind) branch, satisfies
\begin{align}
 \Prb(\FAIL\mid\text{ordinary})
 &\le\Prb(S_{j-1}(v)\ge\delta),\label{iid:low-sandwich}\\
 \Prb(\PASS\mid\text{ordinary})
 &\le\Prb\left(\sum_{b=1}^{K_j/2}Y_{j-1,b}(v)<2\delta\right).
 \label{iid:high-sandwich}
\end{align}
At level one, $1-q_1(v)=(1-\theta)d(v)/m$, where $d(v)$
is the degree of $v$.

\end{lemma}
\begin{proof}

A neighbor whose first failed test is $i$ invokes total budget
$\sum_{a=1}^iK_a=H_{i-1}$, exactly the contribution subtracted from
the residual.  Each guard-limited classification has total permitted
budget below $M_j=\delta K_j$.  Before the last update or a prefix
kill, completed contributions total less than the initial threshold
$U_j<2M_j$.  Adding the last classification, including its hypothetical
completion if necessary, costs less than another $M_j$.  Level-one
and blind-PASS calls have no children.
A residual crossing has raw score at least $U_j/K_j\ge\delta$;
a prefix kill has one hypothetical full mark at least $\delta$.
There are at most $K_j$ own sources, proving~\eqref{iid:low-sandwich}.
On an ordinary PASS, completed score is below $2\delta K_j$.
Each child occupies at most twice its budget, so children use fewer
than $6\delta K_j$ positions of the parent's RUN.  At most one own
source has an unfinished classification.  Since $\delta\le1/24$
and $K_j\ge4$, the first $K_j/2$ comparison marks are complete and
have sum below $2\delta$.  This proves~\eqref{iid:high-sandwich}.
The level-one identity follows from the immediate incidence guard.
\end{proof}
\subsection{Low loads survive; high loads disappear}

The low-load estimate needs only a second moment. The useful split is
between the inherited contribution of a comparison mark and its integer
top-level contribution. The latter is paid for directly by its mean;
the inherited variances form a geometric sum across recursion levels.

\begin{lemma}[Survival bounds]\label{iid:survival}
For every vertex $v\in V$ and integer $r\ge0$,
\begin{equation}
 1-p_{r+1}(v)\le2a_r(v)
 \quad\text{if }a_r(v)\le\delta/6.\label{iid:low-load}
\end{equation}
With $B=1152(c+1)<2^{23}$, for every integer $i\ge0$,
\begin{equation}
 p_i(v)\le\min\{1,(B/a_i(v))^4\},\label{iid:envelope}
\end{equation}
where the right side is one if $a_i(v)=0$.
\end{lemma}
\begin{proof}

Put $\beta_i=c^iD_i(v)/m$, so $a_r=\sum_{i=0}^r\beta_i$.
Decompose $S_i=A_i+B_i$, where $B_i$ is the integer top-level
contribution and $A_i$ contains the lower levels; set $A_0=0$.
Then $\E[B_i]=\beta_i$ and $\Prb(B_i\ge1)\le\beta_i$.
For $i\ge1$, $A_i$ is the average of $c^i$ independent copies of
\[
 X_i=\ind_{\{v\in e\}}\sum_{a<i}c^a
       \ind_{\{L_{e\setminus\{v\}}\ge a\}},\qquad
 \E[A_i]=a_{i-1}.
\]
The geometric prefix satisfies $H_k\le(4/3)c^k$ since $c\ge4$.
Consequently $H_k^2\le2\sum_{a=0}^kc^{2a}$, and expanding the same
survival indicators as in the mean gives
\[
 \operatorname{Var}[A_i]\le\frac{\E[X_i^2]}{c^i}
 \le2\sum_{a<i}c^{a-i}\beta_a=:2R_i.
\]
Assume $a_r\le\delta/6=1/192$. For every $i\le r$, the distance
from $\E[A_i]$ to the threshold $\delta=1/32$ is at least $1/40$.
Chebyshev's inequality therefore gives
$\Prb(A_i\ge\delta)\le3200R_i$.
If $B_i=0$, a threshold crossing by $S_i$ is a threshold crossing
by $A_i$; no independence between these two contributions is needed.
The low sample bound and the level-one incidence identity imply
\[
 \begin{aligned}
 1-p_{r+1}(v)
 &\le\sum_{i=0}^r\Prb(S_i\ge\delta)\\
 &\le a_r+3200\sum_{i=0}^r R_i
 \le\left(1+\frac{3200}{c-1}\right)a_r
 \le2a_r.
 \end{aligned}
\]
Here $R_0=0$, and the penultimate inequality sums the geometric
tails $\sum_{i>a}c^{a-i}\le1/(c-1)$. This proves
\eqref{iid:low-load} without comparing tail probabilities at
different recursion levels.
For $i\ge2$, the sum in~\eqref{iid:high-sandwich} has mean
$a_{i-1}/2$ and summands bounded by two.  If $a_{i-1}\ge8\delta$,
a multiplicative Chernoff bound gives
\[
 q_i(v)\le\theta+e^{-a_{i-1}(v)/32}.
\]
When $a_{i-1}\ge1152$, this is below $(c+1)^{-4}$:
$e^{-36}<2^{-51}$, $\theta<2^{-51}$, and $(c+1)^4<2^{49}$.
Together with $a_i\le(c+1)a_{i-1}$, induction proves
\eqref{iid:envelope}.  For $a_i\le B$ the claim is trivial; above
$B$, the preceding load is at least $1152$ and each step contracts
survival by at least $(c+1)^{-4}$.  The base $i=0$ follows from
$a_0\le1$.
\end{proof}
\subsection{A matching proxy at every cap}

We now turn the survival bounds into a matching estimate. For each
integer cap $J\ge0$, assign edge $uv$ weight
$m^{-1}\sum_{i=0}^J c^ip_i(u)p_i(v)$. Define its total weight and
the weighted degree at a vertex by
\[
 W_J=\frac1m\sum_{\{u,v\}\in E}\sum_{i=0}^Jc^ip_i(u)p_i(v),
 \qquad
 \ell_v^{(J)}=\sum_{i=0}^Jp_i(v)(a_i(v)-a_{i-1}(v)).
\]

Then $\sum_v\ell_v^{(J)}=2W_J$.  If $I$ is the last level passed
by both endpoints of a root with cap $J$, canonical sampling gives
$\Prb(I\ge i\mid\{u,v\})=p_i(u)p_i(v)$ for $uv\in E$
and $0\le i\le J$. For $i>J$ the left side is zero.
Consequently, for a uniform birth edge, its reward
$H_I$ has mean $\mu=W_J$.

\begin{proposition}[Matching proxy]\label{iid:proxy}
For every integer $J\ge0$, the mean reward satisfies
$\mu=W_J\le4B\nu<2^{25}\nu$. If also $h=H_J\le m$, then
\begin{equation}
 \frac{\delta}{64}\frac hm\nu
 \le\mu=W_J\le4B\nu<2^{25}\nu.\label{iid:proxy-bound}
\end{equation}
\end{proposition}
\begin{proof}

The decreasing envelope in~\eqref{iid:envelope} has integral at most
$2B$.  Its right-endpoint Riemann sum therefore gives
$\ell_v^{(J)}\le2B$.  The endpoints of a maximum matching form a
vertex cover, so $W_J\le4B\nu$.
For the lower bound, let $\gamma=\delta/8$ and
$C_J=\{v:p_J(v)\le1/2\}$.  For $J\ge1$, every $v\in C_J$ has
$a_{J-1}(v)\ge\gamma$, by~\eqref{iid:low-load}.
At its first crossing $a_j(v)\ge\gamma$, its survival is at least
$1/2$: this is immediate if $j=0$, and otherwise follows from
$a_{j-1}<\gamma<\delta/6$.  Monotonicity of $p_i$ gives
$\ell_v^{(J-1)}\ge a_j(v)/2\ge\gamma/2$.  Hence
\begin{equation}
 |C_J|\le4W_{J-1}/\gamma.\label{iid:crossing-count}
\end{equation}
If $|C_J|\ge\nu/2$, then $W_J\ge\gamma\nu/8
=(\delta/64)\nu$, which suffices since $h\le m$.
Otherwise at least $\nu/2$ edges of a fixed maximum matching avoid
$C_J$.  Each contributes at least $c^J/(4m)$ to the top-level term
of $W_J$.  Since $H_J<2c^J$, this gives
$W_J\ge\nu h/(16m)$, which is stronger than required.
For $J=0$, $W_0=H_0=1$ and $\nu\le m$ prove the claim directly.
\end{proof}
\subsection{Recursive occupation}

We next bound how often each call type occurs in a root. Direct-child
budgets contract, but logarithmic survival weights require a second,
weighted estimate. The matrix below records expected child budget
per unit parent budget.

Let $t_j(u)$ be the expected number of active own-source reads of a
full canonical call  on $(j,u)$, excluding WAIT and padding.  For call
types $x=(i,v)$, $y=(j,u)$, set

\[
 T_{xy}=\ind_{\{i<j,\,uv\in E\}}
       \frac{K_i}{K_j}\frac{t_j(u)}m p_{i-1}(v),
 \qquad
 \omega_i(v)=\frac{K_i}{m}p_{i-1}(v)D_{i-1}(v).
\]
The index set is $\{1,\ldots,J\}\times V$, which is empty
when $J=0$. We set $W_{-1}=0$.  Columns index parents and rows index children. The entry $T_{xy}$
is the expected child budget of type $x$, per unit parent budget of type $y$,
after completing its last unfinished classification up to the guard.
The decision to request an own-source tuple precedes that tuple;
summing these predictable identities proves the displayed formula.
The expected initial budget of top-level endpoint calls, charging
$K_i$ per entry of type $(i,v)$, is dominated coordinatewise by $\omega$,
and $\sum_x\omega_x=2W_{J-1}$.

\begin{lemma}[Occupation bound]\label{iid:occupation}
The matrix $T$ has column sums at most $3\delta$, satisfies
$T\omega\le2\omega$, and has $I-T$ invertible. Put
$g=(I-T)^{-1}\omega$. For a single canonical root, if $N_x$
is the number of actual call entries of type $x=(i,v)$, then
$K_i\E[N_x]\le g_x$. Moreover,
\begin{equation}
 \left\|g/\omega\right\|_{L^2(\omega)}
 \le2\sqrt{2W_{J-1}},\qquad \sum_xg_x\le4W_{J-1}.
 \label{iid:occupation-bounds}
\end{equation}
Here $\|f\|_{L^2(\omega)}^2=\sum_{x:\omega_x>0}\omega_x|f_x|^2$;
zero-weight coordinates are omitted.
\end{lemma}
\begin{proof}

We prove the two matrix inequalities, then sum the expected budget
over recursion depths. Since levels strictly decrease along children,
$T^J=0$ (with the empty-matrix convention at $J=0$), so $I-T$
is invertible. The column bound is the deterministic child-budget bound.
For a parent $(j,u)$, put $r=D_{j-1}(u)/m$.  A hypothetical neighbor
passing all $j-1$ levels ends further own-source reads: either the
prefix guard kills the parent, or this is its final unfinished
classification.  For the ordinary branch,
\[
 t_j^{\rm ord}(u)r\le1-(1-r)^{K_j},\qquad
 1-q_j^{\rm ord}(u)\ge1-(1-r)^{K_j/2}.
\]
The second inequality follows because an ordinary PASS completes the
first $K_j/2$ comparison tuples.  Thus
$t_j(u)r\le2(1-q_j(u))$ after multiplying by $1-\theta$.
Level one satisfies the same bound directly.
Substitution into $T\omega$, followed by
$\sum_{j>i}p_{j-1}(u)(1-q_j(u))\le p_{i-1}(u)$, proves
$T\omega\le2\omega$.
On positive coordinates put $A_{xy}=T_{xy}\omega_y/\omega_x$.
Its row sums are at most two and
$\sum_x\omega_xA_{xy}\le3\delta\omega_y$.  Cauchy--Schwarz,
first in each row and then summed with weights $\omega_x$, gives
$\|A\|_{L^2(\omega)}\le\sqrt{6\delta}<1/2$.
Canonical offspring laws at actual invocations and nonnegative
charges bound expected occupation by
$g=\sum_{k\ge0}T^k\omega$; the sum is finite since child levels
decrease.  Therefore
$g/\omega=\sum_{k\ge0}A^k\mathbf1$ has the asserted norm bound.
A final Cauchy--Schwarz inequality gives the mass bound.
The inequality $T\omega\le2\omega$ also shows that a zero coordinate
cannot be reached from positive coordinates.
\end{proof}
\subsection{Logarithmic charges}

The occupation bound controls unweighted work. To control the
information revealed by rare survival events, we also need a moment
with weight $\log(1/p)$. The next lemma provides this through
$\sqrt p$, after which Cauchy--Schwarz bounds the total charge.
Set $\gamma=\delta/8$, as in the proof of
Proposition~\ref{iid:proxy}.

\begin{lemma}[A square-root load moment]\label{iid:moment}
For every $J\ge0$,
\begin{equation}
 \sum_v\sum_{i=0}^J\sqrt{p_i(v)}(a_i(v)-a_{i-1}(v))
 \le P W_J,\qquad P=8B/\gamma+4<2^{34}.
 \label{iid:sqrt-moment}
\end{equation}
\end{lemma}
\begin{proof}
The square root of the envelope in~\eqref{iid:envelope} has integral
$2B$.  Each vertex therefore contributes at most $2B$.  Over $C_J$,
\eqref{iid:crossing-count} bounds the total by $8BW_{J-1}/\gamma$.
Outside $C_J$, every $p_i(v)>1/2$, so $\sqrt{p_i(v)}\le2p_i(v)$;
that part contributes at most $4W_J$.  If $J=0$, $C_J$ is empty
and the latter argument suffices.
\end{proof}

The information argument uses the following charges.  A root birth pays
$128$.  A call on $(i,v)$ pays 
\[
 128K_i\bigl[1+\log(1/p_{i-1}(v))+\log(1+a_{i-1}(v))\bigr].
\]

A top-level root call pays  an additional $128K_i\log(1/z)$,
where $z$ is its joint entry probability conditional on its root edge:
$z=p_{i-1}(u)p_{i-1}(v)$ for the first endpoint, and
$z=p_i(u)p_{i-1}(v)$ for the second when $u$ is first.
On entering PAD, a root pays
$128H_I[1+\log(1/z_I(e))]$, where $z_i(uv)=p_i(u)p_i(v)$.
These are analytical charges; the streaming algorithm evaluates none
of them.

\begin{proposition}[Expected total root charge]\label{iid:fee}
Fix any integer cap $J\ge0$, and let $\mu=\E[H_I]$ be the
canonical mean reward. Let $\tau\ge1$ be a predictable birth
position: for every $b$, the event $\{\tau=b\}$ is decided before
edge $b$ is read. Conditional on any positive-probability pre-birth
history event (whose intersection with $\{\tau=b\}$ is likewise
decided before that read), the expected total charge of the full,
untruncated padded root born at $\tau$ is at most
\begin{equation}
 K_0\mu,\qquad K_0=2^{28}.\label{iid:fee-bound}
\end{equation}
\end{proposition}
\begin{proof}

Assume $J\ge1$ and write $W=W_{J-1}\le\mu$ and
$\psi_i(v)=\log(1/p_{i-1}(v))$.
Since $p\log^2(1/p)\le3\sqrt p$, Lemma~\ref{iid:moment} gives
\[
 \sum_x\omega_x\psi_x^2\le3PW,\qquad
 \sum_xg_x\psi_x\le2\sqrt{6P}\,W,
\]
where the second inequality uses Lemma~\ref{iid:occupation}.
The envelope also implies
$\log(1+a_{i-1}(v))\le\log(2B)+\psi_i(v)/4$.
Thus all base call charges have expectation at most
\[
 128\bigl[4+4\log(2B)+(5/2)\sqrt{6P}\bigr]W.
\]
For the root surcharge, fix an oriented root edge $uv$ and set
$z=p_{i-1}(u)p_{i-1}(v)$ and $q=q_i(u)$.  The two endpoint
contributions, before the factor $128K_i$, are
$z\log(1/z)+zq\log(1/(zq))$.  Their sum is at most
$2z\log(1/z)+z/e$, since $q\log(1/q)\le1/e$.
Summing over edges and levels, and using Cauchy--Schwarz with
$\sum\omega=2W$, bounds the surcharge by
$128[2\sqrt{6P}+1/e]W$.
Finally, $\Prb(I=i\mid e)\le z_i(e)$ and $H_i<2c^i$ give
\[
 \E\left[H_I\log(1/z_I(e))\right]
 \le\frac2m\sum_e\sum_{i=0}^Jc^iz_i(e)\log(1/z_i(e)).
\]
Let $\eta_i(v)=p_i(v)(a_i(v)-a_{i-1}(v))$ and
$\chi_i(v)=\log(1/p_i(v))$.  The right side is
$2\sum\eta_i(v)\chi_i(v)$.
Here $\sum\eta=2\mu$ and, by Lemma~\ref{iid:moment},
$\sum\eta\chi^2\le3P\mu$.  The expected PAD charge is consequently
at most $128[1+2\sqrt{6P}]\mu$.
Adding the birth charge and using $\mu\ge1$ proves the bound
\[
 128\bigl[6+4\log(2B)+(13/2)\sqrt{6P}+1/e\bigr]\mu
 <2^{28}\mu.
\]
For the last inequality, $P<2^{34}$ and
$(13/2)\sqrt6<255/16$ bound the square-root term, including the
outer factor $128$, by $255\cdot2^{20}=2^{28}-2^{20}$.
The remaining terms are below $2^{16}<2^{20}$, since
$\log(2B)<24$.  At $J=0$, there are no calls and the birth and
PAD charges total $256$, while $\mu=1$.
All estimates are conditional on the pre-birth history because the
fresh root edge is uniform and its canonical laws are phase independent.
\end{proof}
\section{From retained state to a random permutation}
\label{sec:transfer}

We compare one fixed-horizon phase with its execution on independent
uniform edge samples.  Comparing complete transcripts would reveal every
repeated sample and is too expensive.  Instead we compare the state that
determines the next operation.  The exact completion count is used only
after the phase and is excluded from this comparison; a bounded counter
will certify its final value.  Throughout this section, the graph, phase
parameters, and horizon are fixed.
\subsection{An information bound from a forced-label reference}

\paragraph{Probability and information conventions.}
Let $E$ be a finite set of size $m\ge1$, and let
$X_1,\ldots,X_m$ be a uniform permutation of $E$. Set
$A_t=\{X_1,\ldots,X_t\}$ for $0\le t\le m$, with
$A_0=\varnothing$. Let $\Xi$ be a field of private random
variables independent of the permutation, and write $P$ for their
joint law. The field may include future or unused coins; it contains
no input labels. All states are measurable functions of the observed
prefix and this field. All spaces used below are finite or standard
Borel spaces, so the stated conditional distributions exist.

Write $(m)_r=\prod_{k=0}^{r-1}(m-k)$ for the falling factorial,
with $(m)_0=1$. Relative entropy is
$\KL(P\Vert Q)=\int\log(dP/dQ)\,dP$ when $P\ll Q$, interpreted
as an extended real quantity, and is $+\infty$ otherwise.
For a finite-valued variable $V$, let
$\ent_P(V\mid C)=\E_C\bigl[\sum_vP(v\mid C)\log(1/P(v\mid C))\bigr]$.
Mutual information is
$I_P(S;X)=\KL(P_{S,X}\Vert P_S\otimes P_X)$; conditional mutual
information is the corresponding expectation over the conditioning
variable. Zero-mass terms contribute zero.

The next theorem separates the information in explicitly retained
labels from the remaining state. For example, if the state reveals
only one previously seen edge label, its effect on the next edge
costs $\log(m/(m-1))$, rather than the binary length of that label.

\begin{theorem}[Anchored state comparison]\label{tr:thm:anchors}
Fix an integer $0\le t<m$. Let $C_t,V_t$ be measurable functions
of $(X_1,\ldots,X_t,\Xi)$, with $V_t$ finite-valued, and let
$S_t$ be a function of $(C_t,V_t)$. Let
$\mathcal R_t=\mathcal R_t(C_t)\subseteq E$ be a specified label
set, with $r_t=|\mathcal R_t|$ and
$\mathcal R_t\subseteq A_t$ almost surely under $P$.
Suppose a probability law $Q_t$ on $(A_t,C_t)$ satisfies
\[
 Q_t(A_t\mid C_t)=
 \Unif\{A\subseteq E:|A|=t,\ \mathcal R_t\subseteq A\}
 \quad(Q_t\text{-almost surely}),
\]
and, for finite real bounds $d_t,s_t$,
$\KL(P_{A_t,C_t}\Vert Q_t)\le d_t$ and
$\ent_P(V_t\mid C_t)\le s_t$. Then
\begin{equation}\label{tr:eq:generic-information}
 I_P(S_t;X_{t+1})
 \le \frac{d_t+s_t}{m-t}
       +\E_P\!\left[\log\frac{m}{m-r_t}\right].
\end{equation}
Moreover, let $0\le T\le m$ be an integer. Suppose the permutation
and independent-sample state processes have the same initial law and
the same possibly time-dependent transition kernels $K_t$: under
$P$, the conditional law of $S_{t+1}$ given $(S_t,X_{t+1})$ is
$K_t$, and the independent-sample law satisfies
$R_{t+1}=(R_t\otimes\Unif E)K_t$. Writing $P_t=\calL_P(S_t)$,
\begin{equation}\label{tr:eq:state-induction}
 \KL(P_T\Vert R_T)
 \le \sum_{t=0}^{T-1} I_P(S_t;X_{t+1}).
\end{equation}
The references $Q_t$ need not be consistent at different times.
\end{theorem}
\begin{proof}

We first add the auxiliary variable to the reference, then apply
subset entropy contraction to the unread edges. Finally we sum the
one-step inequalities for the common transition kernels.
Extend the reference by
\[
 Q'_t(A,C,V)=Q_t(A,C)P(V\mid C).
\]
The chain rule  gives
\[
 \KL(P_{A,C,V}\Vert Q'_t)
 =\KL(P_{A,C}\Vert Q_t)+I_P(A;V\mid C)
 \le d_t+s_t.
\]

Under this reference the conditional law of \(A\), given \(C,V\), remains
the uniform forced-label slice.
We use the following contraction. Let $1\le b\le M$ be integers
and write $U_b=\Unif\binom{[M]}b$. If \(\xi\) is a law on
the $b$-subsets of $[M]$ and the kernel $K$ selects a uniform
member of its input set, then
\begin{equation}\label{tr:eq:slice}
 \KL(\xi K\Vert\Unif[M])
 \le \frac1b\KL\left(\xi\middle\Vert
                       \Unif\binom{[M]}b\right).
\end{equation}

Indeed, for arbitrary real numbers \(f_x\), the variational inequality
and Maclaurin's inequality give

\[
 \KL(\xi\Vert U_b)
 \ge \E_\xi\bigl[\sum_{x\in B}f_x\bigr]
       -\log\E_{U_b}\!\left[\exp\left(\sum_{x\in B}f_x\right)\right]
 \ge b\left(\E_{\xi K}[f]
       -\log\left(\frac1M\sum_x e^{f_x}\right)\right).
\]
Here Maclaurin's inequality bounds the average of
$\prod_{x\in B}e^{f_x}$ over $b$-subsets by
$(M^{-1}\sum_xe^{f_x})^b$, and uniform selection gives
$\E_\xi[\sum_{x\in B}f_x]=b\E_{\xi K}[f]$.
Taking the supremum over \(f\) proves \eqref{tr:eq:slice}.
Nonanticipation and independence of \(\Xi\)  imply that, under the actual
law, \(X_{t+1}\) is uniform on \(E\setminus A_t\) conditional on
\((A_t,C_t,V_t)\).  Apply \eqref{tr:eq:slice} conditionally to this
complement, with \(b=m-t\) and ground set \(E\setminus\mathcal R_t\).
The conditional part of the preceding KL bound yields

\[
 \E\!\left[\KL\bigl(P_{X_{t+1}\mid C_t,V_t}
                     \Vert\Unif(E\setminus\mathcal R_t)\bigr)\right]
 \le \frac{d_t+s_t}{m-t}.
\]
For any law \(\zeta\) supported outside \(R\),

\[
 \KL(\zeta\Vert\Unif E)
 =\KL(\zeta\Vert\Unif(E\setminus R))
     +\log\frac m{m-|R|}.
\]
Since $S_t$ is a function of $(C_t,V_t)$ and the next label is
marginally uniform on $E$, data processing proves
\eqref{tr:eq:generic-information}.

Finally, under independent sampling the state/next-edge pair has
law $R_t\otimes\Unif E$. Under the permutation law its next
edge is marginally uniform but may depend on the retained state.
Data processing through the common transition kernel and the chain
rule therefore give
\[
 \KL(P_{t+1}\Vert R_{t+1})
 \le \KL(P_{S_t,X_{t+1}}\Vert R_t\otimes\Unif E)
 =\KL(P_t\Vert R_t)+I_P(S_t;X_{t+1}).
\]
Summation proves \eqref{tr:eq:state-induction}.
\end{proof}
The reference conditions in this theorem are useful because a retained
edge can be represented by the position where it was read.  Its label
then costs one forced element of the prefix, rather than its full binary
description.  The next lemma gives the corresponding likelihood
calculation.

\begin{lemma}[Descriptor likelihood]\label{tr:lem:likelihood}

Fix $0\le t<m$ and write $\Xi=(\Xi_i)_{i\in\mathcal I}$
as a product of coordinate measurable spaces. Let \(\Gamma\) be
a finite-valued descriptor determined measurably by
$(X_1,\ldots,X_t,\Xi)$.  A descriptor \(\gamma\) prescribes $r_\gamma\ge0$ distinct
labels in $E$ at distinct positions in \([t]\), together with a
measurable private-coin event \(B_\gamma\) of probability
$b_\gamma=P(\Xi\in B_\gamma)>0$. Every actual descriptor
satisfies these prescriptions almost surely. Choose a proposal law
$q_\gamma>0$ with $\sum_\gamma q_\gamma=1$, and define \(Q^0\) by
sampling \(\gamma\), a uniform permutation conditional on its prescribed
positions, and independently sampling \(\Xi\) conditional on \(B_\gamma\).
For each \(\gamma\), omit a fixed coordinate set
$I_\gamma\subseteq\mathcal I$ such that membership in $B_\gamma$
depends only on $\Xi|_{I_\gamma^c}$, and let
\(\Omega=\Xi|_{I_\Gamma^c}\).
Suppose a measurable projection
$Y=Y(\gamma,X_1,\ldots,X_m,\Xi)$ reproduces a measurable
state variable $Y_{\rm state}$ almost surely on actual support and,
under \(Q^0\), its conditional law given
\((\Gamma=\gamma,X_1,\ldots,X_m,\Omega)\) is dominated by
\(K_\gamma\rho_\gamma\), where $K_\gamma\ge1$ and
$\rho_\gamma$ is a probability law depending only on \(\gamma\).
Domination means an inequality of conditional measures for every
measurable output set, almost surely under $Q^0$.
Replace this conditional law by \(\rho_\gamma\), independently of the
permutation and \(\Omega\).  The resulting law \(Q\), with
\(C=(\Gamma,\Omega,Y_{\rm state})\) under $P$, has a
uniform forced-label slice and satisfies
\begin{equation}\label{tr:eq:descriptor-kl}
 \KL(P_{A_t,C}\Vert Q_{A_t,C})
 \le \E_P\left[
      -\log q_\Gamma-\log(m)_{r_\Gamma}
      +\log b_\Gamma+\log K_\Gamma\right].
\end{equation}
\end{lemma}
\begin{proof}
Relative to counting measure on descriptors times the base
permutation/coin law, the actual augmented density is the indicator that
the descriptor is the actual one.  On its prescribed support the
reference density is \(q_\gamma(m)_{r_\gamma}/b_\gamma\).
Thus the KL divergence before projection is exactly the expectation of
the first three terms in \eqref{tr:eq:descriptor-kl}.
Data processing and the stated domination add at most
\(\E[\log K_\Gamma]\): write the log likelihood ratio through the
pushforward of \(Q^0\) and bound the additional density ratio.
After replacement, conditional on \(C\), the only restriction on the
permutation is its prescribed positions.  Consequently its prefix set
is uniform among the \(t\)-subsets containing the prescribed labels.
No conditioning on whether the proposed descriptor is an executable
global state is imposed on the reference.
\end{proof}
\subsection{The joint reference for one sampler}

We now specify a proposal law for the actual stack. Its weights
are chosen so that the uniform-edge factors cancel the cost of
forcing anchor labels, while the surviving logarithms are bounded
by the charges from Section~\ref{sec:iid}.

Fix a nonempty simple graph $G=([n],E)$, integers
$0\le t<T<m/2$, $L\ge1$, and $J\ge0$, and a real $w>0$.
Use the recursive-test parameters of Section~\ref{sec:parameters},
write $\mu=W_J$, and set the birth rate to $\lambda=1/L$.
Observation is immediately after all returns and arithmetic at boundary
\(t\), and before the next source read.  Besides \(\IDLE\) and \(\Dead\),
there are two kinds of descriptors.
A \emph{call descriptor} is the chain of source anchors 
\[
 \gamma=((b_0,e_0),\ldots,(b_{r-1},e_{r-1})),
 \qquad r\ge1,\quad 1\le b_0<\cdots<b_{r-1}\le t,
\]
from the root edge to the youngest current call.
The labels $e_0,\ldots,e_{r-1}\in E$ are distinct. The root
endpoints are ordered by their
names.  Starting from boundary \(b_0\), the all-\(\PASS\) calendar for
the two root endpoints determines the current level, endpoint role,
invocation boundary $\sigma_0$, and calendar $(a_0,d_0)$;
a valid root slot has $1\le j_0\le J$ and
$\sigma_0\le t<d_0$. Similarly, an anchor
\(e_k=\{v_{k-1},v_k\}\) at a position
$a_{k-1}<b_k\le d_{k-1}$ in its parent's RUN determines the
all-\(\PASS\) calendar of the classification of \(v_k\).
Only guarded child levels $1\le i<j$ at a level-$j$ parent
are permitted.  A proposed current child must satisfy
\(\sigma_{\rm child}\le t<D_{\rm child}\), and \(t\) must precede every
ancestor deadline.  Its own return may lie after an ancestor's deadline:
such a child is still in WAIT and will be canceled before RUN.
Levels strictly decrease down the chain.  These tags are deterministic
functions of the anchors and \(t\).
Give \(\IDLE\) weight \(1\) and \(\Dead\) weight \(w>0\).
A root call option on edge \(uv\), with \(u\) tested first and current
level \(j\), has weight \(\lambda z/m\), where
\[
 z=p_{j-1}(u)p_{j-1}(v)
 \quad\text{or}\quad
 z=p_j(u)p_{j-1}(v),
\]
according as the first or second endpoint is current.
At a current level-\(j\) frame on \(v\), after \(k\) anchors have been
chosen, a permitted child \((b,\{v,u\},i)\) has weight
\(p_{i-1}(u)/(m-k)\).  Let $Z$ be the total weight of permitted children. Give $\NULL$
weight \(1\), normalize these choices by \(\mathcal N=1+Z\), and either continue with the chosen
child or end at NULL.  The new source label must differ from all earlier
anchors.  This finite tree of choices is a probability distribution;
it need not reproduce a global execution on the subsequently sampled
input.
A \emph{padding descriptor} is
\((\PAD,b,e,i)\), with $e\in E$, where
\[
 1\le b\le t<b-1+7H_i,\qquad 0\le i\le J.
\]
Its weight is \(\lambda z_i(e)/m\), with
\(z_i(uv)=p_i(u)p_i(v)\).  It has the single anchor \(X_b=e\).
No condition that a root actually finishes at level \(i\) is imposed.
Let \(Z_0(t)\) normalize all root, padding, idle, and dead options.
\begin{lemma}[Proposal normalizers]\label{tr:lem:normalizers}
The preceding reference satisfies
\begin{equation}\label{tr:eq:normalizers}
 Z_0(t)\le 1+w+128\mu/L,
 \qquad
 \mathcal N\le 1+64a_{j-1}(v)
\end{equation}
at every level-$j$ frame on $v$ along a valid descriptor at
boundary $t$, including its terminal $\NULL$ normalizer.
\end{lemma}
\begin{proof}
A neighbor classification currently at level \(i\) was born within
\(2\sum_{\ell=1}^iK_\ell=2H_{i-1}<3K_i\) positions.
There are at most \(4K_i\) possible integer births.
The two-endpoint root calendar similarly gives at most \(6K_j\)
births for either endpoint's current level-\(j\) tag.  Since the
second endpoint's prerequisite is at most the first one's,
the total call-root weight is at most
\[
 \frac{12\lambda}{m}
 \sum_{uv\in E}\sum_{j=1}^J
       K_j p_{j-1}(u)p_{j-1}(v)
 \le 64\lambda\mu.
\]
For a child normalizer, \(k\le t<m/2\), and therefore
\[
 Z\le\frac4{m-k}\sum_{i<j}K_iD_{i-1}(v)
 \le 8a_{j-1}(v)\le64a_{j-1}(v).
\]
Guards, distinct-label requirements, and deadlines only remove options.
There are fewer than \(8H_i\) padding births for a fixed \(e,i\).
Since \(H_i<2c^i\), their total weight is at most
\[
 \frac{8\lambda}{m}\sum_e\sum_{i=0}^JH_i z_i(e)
 \le16\lambda\mu.
\]
Adding the mode weights proves \eqref{tr:eq:normalizers}.
\end{proof}
For an active call chain let \(p_k\) be the prerequisite
\(p_{j_k-1}(v_k)\) of child \(k\), and let
\(\mathcal N_0,\ldots,\mathcal N_{r-1}\) be its frame normalizers,
including the terminal NULL.  Its proposal probability is
\[
 q_\gamma=\frac{\lambda z}{mZ_0}
     \prod_{k=1}^{r-1}
       \frac{p_k}{(m-k)\mathcal N_{k-1}}
     \frac1{\mathcal N_{r-1}}.
\]
Force the anchors and only the root-arrival gate at \(b_0\); the latter
event has probability \(\lambda\).
The factors \((m)_r\) and \(\lambda\) in
Lemma~\ref{tr:lem:likelihood} cancel exactly.  The resulting log
likelihood is
\begin{equation}\label{tr:eq:call-likelihood}
 \log Z_0+\log(1/z)+
 \sum_{k=0}^{r-1}\log\mathcal N_k+
 \sum_{k=1}^{r-1}\log(1/p_k).
\end{equation}
For padding it is \(\log Z_0+\log(1/z_i(e))\); for idle it is
\(\log Z_0\); for dead it is \(\log Z_0+\log(1/w)\).
Thus birth positions, labels, and calendar tags incur no separate
description-length term.
\subsection{Residual channels and state reconstruction}

The likelihood calculation alone does not account for the numerical
residuals.  Their random initial thresholds supply a constant-cost
channel.
For integers $M\ge1$ and $s\ge0$, if \(U\) is uniform on
\(\{M,\ldots,2M-1\}\),
the law of \(U-s\), replaced by \(\bot\) when nonpositive, is dominated
by three times
\[
 \rho_M=\tfrac13\delta_\bot+
            \tfrac23\Unif\{1,\ldots,2M\}.
\]
Indeed, each positive value has probability at most \(1/M\), and the
atom at \(\bot\) has probability at most \(1\).  The same domination
holds if a threshold-independent stopping rule forces \(\bot\).
To apply this observation simultaneously to the stack, assign private
primitives to all potential addresses
\[
 (\text{root birth},\text{classification source birth},
                 \text{endpoint role},\text{level},\text{operation kind}).
\]
A top-level endpoint call uses classification source birth zero;
a classification started at physical position $b\ge1$ uses birth
$b$. The operation kind distinguishes thresholds, blind branches, and monitor
gates; root-arrival, root-birth-monitor, padding-monitor, and auxiliary
tester coins have separate families.  Phase and copy identifiers are
included when necessary.  A physical source edge starts at most one
classification, which tests each level at most once, so actual calls
always use fresh coordinates.
For a call descriptor \(\gamma\), omit exactly the threshold coordinates
of its \(r\) current calls, and expose all other private primitives as
\(\Omega\).  The omitted-coordinate mask is fixed by \(\gamma\).
For each current call $k\in\{0,\ldots,r-1\}$, run an
isolated, unmonitored \emph{ghost} from
its reconstructed invocation boundary, suppressing its own threshold
check.  Preserve the blind branch, guards, all subquery rules, and
fixed deadlines.  For $k<r-1$, the ancestor ghost stops at boundary $b_{k+1}-1$,
immediately before the source edge of its current child; the
youngest ghost, $k=r-1$, stops at \(t\).
Every ghost stops all recursive activity at its freeze or observation
boundary and reads no later input edge. The hypothetical i.i.d. continuation
used for the sample sandwich is never used here.
Its completed contributions give a nonnegative score \(s_k\).
If a guard or blind branch has already decided the call, force the
residual to \(\bot\); otherwise return \(U_k-s_k\) when positive,
and \(\bot\) when nonpositive.  Retain also its ordinary/blind and live/decided flag, marking
the call decided exactly when the resulting residual is $\bot$.
These ghosts read none of the omitted thresholds.  An ancestor stops
before its current child's source birth, so all subcalls it can create
have earlier source addresses than every distinguished descendant.
A younger current call invokes only smaller levels and cannot read a
distinguished ancestor's threshold.  Its own threshold is suppressed.
This argument uses only the proposed chronology and levels, and
therefore also holds for reference descriptors that are not actual
executions.
On actual support, an ancestor is still undecided and its pending
neighbor has contributed nothing; its ghost score at the freeze is
exactly its actual score.  The youngest ghost agrees with the actual
call until a decision.  After threshold failure its ghost score cannot
decrease below that threshold; a later guard keeps the forced
\(\bot\) value.  Blind success has its cleared flag.  Ordinary success
occurs only at the fixed return boundary, where the frame is removed.
Consequently the projection reproduces the actual residuals and flags,
including early decisions that are still being padded.
Under the forced reference, conditional on the descriptor, full
permutation, and \(\Omega\), the ghost scores are fixed and the omitted
thresholds are independent.  No such independence is asserted under
the actual law.
The joint residual channel is dominated by \(3^r\) times the product
of the \(\rho_{M_{j_k}}\).  A uniform reference on four possible flags
per frame adds at most \(4^r\).  Replacing the whole channel therefore
costs at most \(r\log12\).
The replacement is independent of the permutation and exposed
primitives.  Lemma~\ref{tr:lem:likelihood} now applies.
In padding, idle, and dead modes the threshold mask is empty, and no
ghost is needed.
For clarity, the compared state contains the following fields.
\begin{center}
\begin{tabular}{@{}p{0.18\linewidth}p{0.73\linewidth}@{}}
\toprule
Record & Retained information\\
\midrule
Root & Its source birth and endpoints, and the current endpoint and level.\\
Current call & Its vertex, level, calendar, residual or cleared value,
and a constant-size flag.\\
Pending neighbor & Its source birth, vertex, and current classification level.\\
Padding & The root birth, edge, terminal level \(i\), and deadline
\(b-1+7H_i\).\\
Auxiliary state & The monitor bit and a finite counter \(V_t\).\\
\bottomrule
\end{tabular}
\end{center}
The anchors and their deterministic tags reconstruct all label and
calendar fields; the projection supplies residuals and flags.
The terminal level in padding is included because it determines the
next completion.  Finished-neighbor records, popped frames, causes of
decisions, and transition scratch are erased before the next read.
Only the residual is retained, never the initial threshold and
accumulated score separately.
The exact completion count is omitted only because no within-phase
transition consults it.  Every field used by the auxiliary counter's
next update is included.
\subsection{Prepaid occupation and the comparison inequality}

Fix the same graph and phase parameters as above, and any real
monitor scale for the construction below. Let $V_t$ be any finite-valued
measurable function of the prefix and private coins adjoined to the
compared state. Write $C_t=(\Gamma_t,\Omega_t,Y_t)$ for the
descriptor, exposed private primitives, and residual/flag vector.
For a live call descriptor define
\[
 \calL_t=\log(1/z)+
   \sum_{k=0}^{r_t-1}\log\mathcal N_k+
   \sum_{k=1}^{r_t-1}\log(1/p_k)+r_t\log12.
\]
In padding put \(r_t=1\) and
\(\calL_t=\log(1/z_i(e))+\log12\).
Set both quantities to zero in idle and dead modes.
Under the random-permutation law, the preceding construction and
Theorem~\ref{tr:thm:anchors} give
\begin{align}
 I_{\WOR}(S_t;X_{t+1})
 &\le
 \frac{\log(1+w+128\mu/L)
       +\Prb(\Dead_t)\log(1/w)+\E[\calL_t]
       +\ent(V_t\mid C_t)}{m-t}
 \notag\\[-2pt]
 &\hspace{2em}+\E\!\left[\log\frac m{m-r_t}\right].
 \label{tr:eq:predictive}
\end{align}
We next relate the live charge to the causal charges from the i.i.d.
analysis.  Charge \(128\) at root birth.  At entry to a level-\(i\)
call on \(v\), charge
\[
 128K_i\left[1+\log\frac1{p_{i-1}(v)}
                     +\log(1+a_{i-1}(v))\right].
\]
A top-level endpoint entry has the additional charge
\(128K_i\log(1/z)\), with the prerequisite \(z\) for that endpoint.
At entry to padding, charge
\(128H_I[1+\log(1/z_I(e))]\).
All quantities in these charges are fixed functions of the graph and
the current type; the executable estimator does not evaluate them.
\begin{lemma}[Pathwise coverage]\label{tr:lem:coverage}
For every integer cap $J\ge0$, external horizon $0\le T<m/2$,
and monitor scale, use the recursive-test parameters of
Section~\ref{sec:parameters}. If each entry charge is accepted before
its state is created, then on every sample path the total
accepted charge \(F_{\rm acc}\), up to any external horizon or monitor
death, satisfies
\begin{equation}\label{tr:eq:coverage}
 \sum_{0\le t<T}(\calL_t+r_t)\le F_{\rm acc}.
\end{equation}
It also bounds the number of observation boundaries $0\le t<T$
at which the sampler is busy and the monitor is alive.
\end{lemma}
\begin{proof}
A level-\(i\) frame persists for at most \(2K_i\) positions.
By \eqref{tr:eq:normalizers},
\(\log\mathcal N\le\log64+\log(1+a_{i-1}(v))\).
Its prerequisite logarithm, residual/flag cost \(\log12\), and one
anchor unit are therefore covered by its entry charge.
The root prerequisite is covered by the top-level surcharge.
The root-birth charge covers the birth convention.
In padding there are at most \(7H_I\) observation positions, and,
writing \(x=\log(1/z_I(e))\ge0\),
\[
 7H_I(x+\log12+1)\le128H_I(1+x).
\]
Cancellation or an external horizon only shortens these intervals.
A rejected charge creates no live state.  Since \(r_t\ge1\) whenever the
sampler is busy and alive, the same bound covers its busy area.
\end{proof}
For $B_{\rm mon}>0$ and an integer $T\ge1$, use monitor
scale $B_{\rm mon}T$. Accept each nonnegative causal charge $x\ge0$ with a fresh
independent probability \(e^{-x/(B_{\rm mon}T)}\); on rejection erase the core
and auxiliary counter and enter the absorbing state \(\Dead\).
Combine a call's base charge and top-level surcharge before using its
single entry gate.  The monitor keeps no accumulated charge.
For every input edge-stream law independent of the private coins,
\begin{equation}\label{tr:eq:monitor}
 \E[F_{\rm acc}]
 \le B_{\rm mon}T\,\Prb(\Dead_T)\le B_{\rm mon}T,
\end{equation}
because \(xe^{-x/(B_{\rm mon}T)}\le B_{\rm mon}T(1-e^{-x/(B_{\rm mon}T)})\) and the first
rejection events are mutually exclusive.
The compared state has a common causal update kernel under both source
laws.  Calendar fields determine source ownership and returns; each
new private coordinate is fresh conditional on the actual past.
Charges depend only on the current type and fixed graph parameters, and
the monitor needs only its alive/dead bit.  The same holds for any
bounded auxiliary counter updated from visible completion events and
fresh coins.  The exposed field in a one-time certificate is not part
of this update filtration.
\begin{corollary}[Fixed-phase comparison]\label{tr:cor:phase}
Fix a simple graph with $m>0$ edges, integers $1\le T<m/2$
and $J\ge0$, a power-of-two rate $L\ge1$, and $B_{\rm mon}>0$.
Use the recursive-test constants of Section~\ref{sec:parameters}
and write $\mu=W_J$. Let the finite-valued auxiliary counter start
in a fixed state and reset to a fixed state on $\Dead$. While
alive, its next value is a measurable function of the step position,
the pre- and post-step retained core (including any root completion),
its own value, and a fresh independent coin. Suppose
$\ent_{\WOR}(V_t\mid C_t)\le s$ for every integer $0\le t<T$.
For the monitored sampler at scale $B_{\rm mon}T$, with $w=1$,
its terminal compared-state laws $P_T$ under random order and $R_T$
under independent samples satisfy, in nats,
\begin{equation}\label{tr:eq:phase-kl}
 \KL(P_T\Vert R_T)
 \le
 \frac{T[\log(2+128\mu/L)+s]+\E_{\WOR}[F_{\rm acc}]}{m-T}.
\end{equation}
\end{corollary}
\begin{proof}
Since \(r_t\le t\),
\(\log(m/(m-r_t))\le r_t/(m-t)\).
Sum \eqref{tr:eq:predictive}, apply
\eqref{tr:eq:state-induction}, and use
\eqref{tr:eq:coverage}.
\end{proof}
The i.i.d. analysis bounds the expected full charge of a newly born root by
\(K_0\mu\), where \(K_0=2^{28}\).
Together with its exact duration \(7H\), this bounds the monitor's
i.i.d. death probability.  Corollary~\ref{tr:cor:phase} then transfers a
bounded-counter certificate to the random permutation.
For sequential phases, the argument is applied anew conditional on
the complete actual past at each phase's start.  This fixes its
residual graph and parameters while leaving the unread suffix uniformly
ordered; it does not condition on a future matching-survival event.
\section{Estimating from a fixed phase}
\label{sec:phase}

Fix a nonempty residual graph with $m$ edges. A phase has dyadic
rate $L\ge1$, horizon $T$, and cap $h=H_J\le\kappa T$.
Write $\mu=\E_{\IID}[H_I]$. The parameters $\kappa$, the monitor
constant $B_{\rm mon}$, the tester constants, and the normalization
$S$ will be fixed in Section~\ref{sec:parameters}.
\subsection{An i.i.d. completion count}

We first analyze the completion count on an infinite independent edge
stream, continuing past the external horizon. This avoids conditioning
the final, possibly unfinished root on completing before $T$.
Set $D_0=0$. For each $i\ge1$, let $G_i$ be the idle waiting
time after $D_{i-1}$, including the birth position of root $i$.
Then $G_i$ is geometric on $\{1,2,\ldots\}$ with mean $L$,
the root is born at $b_i=D_{i-1}+G_i$, and its full reward at
cap $J$ is $Q_i=H_{I_i}$. Its padded completion boundary and
cycle length are
\[
 D_i=b_i-1+7Q_i,\qquad R_i=D_i-D_{i-1}=G_i-1+7Q_i.
\]
Define $Y_i=G_i+Q_i$ and $y=L+\mu$. For every $i\ge1$,
\begin{equation}
 Y_i\le R_i\le7Y_i.\label{phase:cycles}
\end{equation}
Under the independent-sample law, conditional on the stopped history
through $D_{i-1}=\sum_{j<i}R_j$, almost surely,
\begin{equation}
 \E[Y_i]=y,\qquad \operatorname{Var}[Y_i]\le(L+h)y.
 \label{phase:moments}
\end{equation}
Indeed, conditional on the idle wait and all its discarded edges,
but before the root edge is revealed, the root still has its
canonical mean $\mu$. This gives zero conditional covariance of
$G_i$ and $Q_i$, while $\operatorname{Var}[G_i]=L(L-1)$ and
$Q_i^2\le hQ_i$. In particular, for deterministic $q$,
\[
 \E\bigl[\sum_{i=1}^qY_i\bigr]=qy,\qquad
 \operatorname{Var}\Bigl[\sum_{i=1}^qY_i\Bigr]\le q(L+h)y.
\]
This uses orthogonality of the centered increments; no independence
of stopped or completed rewards is needed.
\begin{lemma}[i.i.d. count]\label{phase:iid-count}
Let $L\ge1$ be dyadic and $T\ge1$ an integer. Let $J$
be the cap chosen by $\proc{Phase}(L,T)$ and $h=H_J$.
Suppose $L,h\le\kappa T$. Under the joint law of independent
uniform edges and independent phase coins, for the constants
$\ell,u,\kappa$ fixed in Section~\ref{sec:parameters},
\[
 \Prb_{\IID}\{\ell T/y\le N_T\le uT/y\}
 \ge1-2^{-10}.
\]
\end{lemma}
\begin{proof}
Put $q_-=\lfloor T/(14y)\rfloor$ and $q_+=\lceil4T/y\rceil$.
Since $y\le2\kappa T$ and $\kappa\le2^{-15}$,
$q_-\ge T/(28y)\ge\ell T/y$.
If $N_T<q_-$, the first $q_-$ cycles exceed $T$, so
$\sum_{i\le q_-}Y_i>T/7$. Its mean is at most $T/14$ and
its variance at most $T(L+h)/14$. Chebyshev gives failure at
most $14(L+h)/T\le28\kappa$.
If $N_T\ge q_+$, then $\sum_{i\le q_+}Y_i\le T$, despite
mean at least $4T$. Chebyshev bounds this probability by
$16(L+h)/(9q_+y)\le8\kappa/9$.
Otherwise $N_T<4T/y\le uT/y$. The two failures sum to less
than $32\kappa\le2^{-10}$. An unfinished final cycle is included
in these implications.
\end{proof}
\subsection{A bounded tester for an output-only count}

Suppose $1\le L,h\le\kappa T$, and put $g=T/(L+\mu)$.
The exact count $N_T$ is not part of the
compared state. Instead, for analysis only, independently retain each
completion with probability $p=b_{\rm test}/g$ and keep its count
$V$, initialized at $V_0=0$ and capped at $C_{\rm test}=4u b_{\rm test}+1$.
Use an independent coin addressed by each physical boundary, using
it only when a root completes there. The sampler and monitor do
not inspect these coins. At monitor death reset $V$ to zero and
keep it zero. Accept if
\begin{equation}
 \ell b_{\rm test}/2\le V_T\le2u b_{\rm test}.
 \label{phase:accept}
\end{equation}
The constants give $g\ge2^{14}$, $0<p\le1/8$, and
$\ent(V_t\mid C_t)\le\log(C_{\rm test}+1)<64$.
\begin{lemma}[Source-independent reliability]\label{phase:tester}
Let $m>0$, $1\le L,h\le\kappa T$, and $\zeta=2^{-7}$,
using the constants of Section~\ref{sec:parameters}. Suppose the
tester coins are jointly independent of the input, sampler, monitor,
and conditioning history. Conditional on any sampler/monitor
trajectory on which the monitor survives through $T$, the uncapped
terminal count is $\operatorname{Bin}(N_T,p)$. For each such
trajectory,
\begin{align*}
 \ell g\le N_T\le ug
 &\ \Longrightarrow\ \Prb(\text{accept}\mid\text{trajectory})\ge1-\zeta,\\
 N_T<\ell g/4\ \text{or}\ N_T>4ug
 &\ \Longrightarrow\ \Prb(\text{accept}\mid\text{trajectory})\le\zeta.
\end{align*}
The same conclusions hold conditional on a positive-probability
history event on which the relevant count condition and monitor
survival hold throughout. The conditional binomial identity itself
requires $N_T$ to be constant on that event. These statements hold
for any input source independent of the tester coins.
\end{lemma}
\begin{proof}
Independence gives the binomial law. In the first interval, its mean
lies between $64$ and $2^{14}$; the acceptance interval is
$[32,2^{15}]$. Chernoff bounds give rejection probability at most
$e^{-8}+e^{-2^{14}/3}<2^{-7}$. In the too-small-count case the
mean is below $16$, so the probability of reaching $32$ is at
most $e^{-16/3}<2^{-7}$. In the too-large-count case the mean
exceeds $2^{16}$, and falling to $2^{15}$ has probability at
most $e^{-2^{16}/8}<2^{-7}$. The cap is above the acceptance
interval and changes none of these events.
\end{proof}
The tester's probability depends on $\mu$, but the implemented
algorithm uses only $N_T$ and does not run this tester. The tester
is included solely in the probability comparison. Its boundary-indexed
coins are separate from all call and monitor primitives, so adjoining
them preserves the reference construction in Section~\ref{sec:transfer}.
\subsection{The phase guarantee}

The phase proof combines three events: concentration of the
independent-sample completion count, monitor survival, and tester
acceptance. The state comparison transfers their joint certificate
to random order. We first bound the probability of monitor death.

The monitor accepts a charge $x$ with probability
$e^{-x/(B_{\rm mon}T)}$. The expected unmonitored i.i.d. charge through
$T$ is at most
\begin{equation}
 K_0(T/7+h).\label{phase:iid-charges}
\end{equation}
To verify this for predictably started roots, let $B_T$ be the
number born by $T$. Whether root $i$ is born is determined before
its edge and fresh coins are revealed. Hence
$\mu\E[B_T]=\E\bigl[\sum_{i\le B_T}Q_i\bigr]$. Completed roots have total
reward at most $T/7$, and at most one unfinished root has full
reward at most $h$. The same predictable-birth identity applied to
total root charges, with Proposition~\ref{iid:fee}, proves
\eqref{phase:iid-charges}. Thus i.i.d. monitor death has probability at
most
\begin{equation}
 d_0=\frac{K_0}{B_{\rm mon}}(1/7+\kappa)<3/64.
 \label{phase:death}
\end{equation}
Let $\Lambda(F)=1+\lceil\log_2F\rceil$ for an integer $F\ge1$.

\begin{proposition}[Fixed phase]\label{phase:proposition}
Let $G=([n],E)$ be simple with $m=|E|>0$. Let $F\ge1$
and $T$ be integers, and let $L=2^k$ for an integer $k\ge0$.
Use the constants of Section~\ref{sec:parameters}.
Suppose $\nu(G)/L\le F$ and
\begin{equation}
 T\ge2^{32},\qquad L<\kappa T,\qquad
 T\le a m/\Lambda(F).\label{phase:conditions}
\end{equation}
Run Algorithm~\ref{alg:phase} on the first $T$ edges of a
uniform permutation of $E$. With probability at least $9/10$
over the permutation and phase coins,
$N_T>0$ and
\begin{equation}
 2^{-5}(L+\mu)\le T/N_T\le2^7(L+\mu).
 \label{phase:count}
\end{equation}
Moreover, the independent-sample mean reward $\mu$ at its chosen
cap satisfies
\begin{equation}
 2^{-39}(T/m)\nu(G)\le\mu<2^{25}\nu(G).
 \label{phase:proxy}
\end{equation}
\end{proposition}
\begin{proof}
For the parameter choice in Section~\ref{sec:parameters},
$\kappa T\ge2^{17}>H_1$. Maximality of $h=H_J$ and
$H_{J+1}=1+ch\le2ch$ give $h>\kappa T/(2c)$.
Proposition~\ref{iid:proxy} proves~\eqref{phase:proxy}, since
$\delta\kappa/(128c)=2^{-39}$ and $h<m$.
Adjoin the bounded tester to the monitored core. Its entropy is
below $64$, and $\mu/L<2^{25}F$ gives
$\log(2+128\mu/L)\le64\Lambda(F)$.
Corollary~\ref{tr:cor:phase} and the accepted-charge bound give
\[
 \KL(P_T\Vert R_T)
 \le\frac{T}{m-T}\,[B_{\rm mon}+64\Lambda(F)+64]
 \le4B_{\rm mon}(T/m)\Lambda(F)
 \le2^{-10}.
\]
Pinsker's inequality therefore gives $\TV(P_T,R_T)<2^{-5}$.
This is a comparison of the terminal monitored core and bounded
tester, after cleanup and before the output-only count is read.
Under i.i.d. input, monitor survival and tester acceptance occur
together with probability at least
$1-2^{-10}-3/64-\zeta$, by Lemmas~\ref{phase:iid-count}
and~\ref{phase:tester}. Transfer this joint event to permutation
input. On monitor survival the monitored and implemented samplers
coincide; tester soundness under the permutation removes at most
another $\zeta$. Consequently
\[
 \ell g/4\le N_T\le4ug
\]
with failure at most
$2^{-10}+3/64+2^{-5}+2\zeta=97/1024<1/10$.
The lower endpoint is positive, and reciprocating the interval gives
\eqref{phase:count}. There is no separate loss for permutation
monitor death: the transferred event already requires survival.
\end{proof}
The next lemma expresses the phase guarantee in the form used by the
outer algorithm. It compares the candidate with the original matching
size, while the phase itself sees only the unread graph.

\begin{lemma}[One refinement step]\label{phase:refinement-step}
Let $G_0$ be a simple graph with $m_0>0$ edges and matching size
$\nu_0\ge1$, in uniformly random edge order. At a phase start,
let $G'\subseteq G_0$ be the unread graph, with $m'$ edges and
matching size $\nu'$. Let $L$ be a dyadic integer, $F\ge1$ an
integer, and $T\ge0$ an integer horizon. Suppose
\begin{equation}
 1\le L\le\nu_0,\qquad \nu_0/L\le F,
 \qquad \nu'\ge\nu_0/2.\label{phase:invariant}
\end{equation}
If $T\ge2^{32}$, $L<\kappa T$, and $T\le am'/\Lambda(F)$,
run $\proc{Phase}(L,T)$ on the next $T$ unread edges with fresh
coins. Its candidate $\lambda$ satisfies
\begin{equation}
 \Prb\{c_*(T/m_0)\nu_0\le\lambda\le\nu_0
       \mid\text{complete phase-start past}\}\ge9/10,
 \qquad c_*=2^{-79}.\label{phase:improvement}
\end{equation}
This conditional statement holds on every positive-probability past
event fixing $G'$ and the phase parameters and satisfying the stated
hypotheses. If instead $L\ge\kappa T$, the candidate $\lambda=L$
satisfies the same interval deterministically.
\end{lemma}
\begin{proof}
Conditioning on the read edges and earlier private coins fixes $G'$
and leaves its unread order uniform. The phase hypotheses therefore
allow Proposition~\ref{phase:proposition} to be applied to $G'$.
On its good count event,
$T/N_T\le2^7(1+2^{25})\nu_0<S\nu_0$, using $L\le\nu_0$
and $\nu'\le\nu_0$. Dividing by $S$, rounding down to a power
of two, and taking the maximum with one preserve the upper bound
$\lambda\le\nu_0$.

For the lower bound, the count and proxy estimates give
$T/N_T\ge2^{-5}\mu\ge2^{-44}(T/m')\nu'$.
Use $\nu'\ge\nu_0/2$, $m'\le m_0$, and the rounding loss
of at most two. Since $1/(2S)=2^{-34}$, the resulting coefficient
is $2^{-5}2^{-39}(1/2)2^{-34}=2^{-79}$.
This proves~\eqref{phase:improvement}.
In the skipped case, $L\le\nu_0$ and
$L\ge\kappa T\ge c_*(T/m_0)\nu_0$, because
$\nu_0\le m_0$ and $c_*\le\kappa$.
\end{proof}

\section{Sequential refinement to a constant}
\label{sec:refinement}

A phase of length proportional to $m/(1+\log F)$ improves an
$F$-factor lower estimate to an $O(1+\log F)$-factor estimate. Iterating this
observation requires budgets that sum to less than a fixed fraction
of the stream and repetition counts whose errors are summable.
We give an explicit schedule so that these requirements also hold
at the last, constant-sized target.
Choose an absolute constant $\epsilon>0$ and integers $R,D\ge1$
and $B_0\ge1$, with $B_0$ a power of two, as in
Section~\ref{sec:parameters}. For each integer $s\ge1$, define
\begin{equation}
 e_s=\left\lfloor\frac{B_{s-1}}{D(s+1)2^s}\right\rfloor,
 \qquad B_s=2^{e_s},\qquad r_s=2R(s+1)-1.
 \label{wrap:targets}
\end{equation}
The choices in Section~\ref{sec:parameters} satisfy
\begin{equation}
 B_s\ge B_{s-1}^2,\quad
 B_{s-1}\ge2^{141\cdot2^{s-1}},\quad
 \Lambda(B_s)\le\frac{2B_{s-1}}{D(s+1)2^s}.
 \label{wrap:growth}
\end{equation}
In particular the sequence is unbounded. Its role is just to specify
targets in reverse: stage $s$ begins with approximation factor at most
$B_s$ and ends with factor at most $B_{s-1}$.
\begin{algorithm}[h]
\caption{$\proc{KnownLength}(m_0)$}
\label{alg:known}
\small
\begin{enumerate}[leftmargin=1.7em,label=\arabic*.,itemsep=3pt]
\item Given an integer $m_0\ge0$, if $m_0=0$, return $0$.
If $1\le m_0\le B_0$, set the
candidate to $1$ and return it at the designated commitment boundary;
ignore the intervening edges and do not execute steps 2--4.
\item Otherwise set $L\gets1$. Generate the targets $B_0,B_1,\ldots$
only until the first index $k$ with $B_k\ge m_0$, using the
safe exponent comparison described below.
\item For $s=k,k-1,\ldots,1$, set
\[
 F_s=\begin{cases}m_0,&s=k,\\ B_s,&s<k,\end{cases}
 \qquad
 T_s=\left\lfloor\frac{\epsilon 2^{-s}m_0}
                         {r_s\Lambda(F_s)}\right\rfloor.
\]
If $L\ge\kappa T_s$, skip this stage. Otherwise run $r_s$ copies of $\proc{Phase}(L,T_s)$ on
consecutive disjoint blocks of $T_s$ edges, with independent private
coins, holding the old $L$ fixed throughout the stage. Set $L$ to the maximum of its old value
and the median candidate.
\item Set the candidate to $L$; ignore the remaining prescribed edges.
Return this candidate at the designated commitment boundary.

\end{enumerate}
\end{algorithm}
To generate targets without storing enormous integers, put
$q=\lceil\log_2m_0\rceil$. After computing $e_s$, if $e_s\ge q$
stop and set $k=s$; otherwise form $B_s=2^{e_s}<m_0$.
Equality is a stopping condition. Thus $B_k$ is never materialized,
and every constructed integer has $O(\log(m_0+2))$ bits.
Growth gives $k=O(\log\log(m_0+2))$ and, more precisely,
\begin{equation}
 2^k<2\log_2m_0/141\quad(m_0>B_0).
 \label{wrap:k-bound}
\end{equation}
For the chosen constants, every nontrivial horizon satisfies
$T_s\ge2^{75}$, and rounding it down loses at most a factor of two.
These deterministic statements hold even on failed executions;
they are verified in Section~\ref{sec:parameters}.
All executed phases consume a prefix of length less than
\begin{equation}
 \sum_{s=1}^k r_sT_s
 \le m_0\sum_{s=1}^k\frac{\epsilon2^{-s}}{\Lambda(F_s)}
 <\epsilon m_0.\label{wrap:prefix}
\end{equation}
At every phase start, therefore, $m'\ge(1-\epsilon)m_0$ and
\[
 T_s\Lambda(F_s)/m'
 \le\frac{\epsilon2^{-s}}{r_s(1-\epsilon)}<a.
\]
The phase condition on $L$ holds exactly when the stage is executed.

The stage invariant is deterministic once its phase candidates are
fixed. The next lemma isolates this step; the theorem that follows
will bound the probability that a stage violates its hypotheses.

\begin{lemma}[A successful stage]\label{wrap:stage}
Use the constants of Section~\ref{sec:parameters}. Let $G_0$ have
$m_0>B_0$ edges and matching size $\nu_0\ge1$, and fix any edge
permutation and private stage coins. Suppose
$1\le L\le\nu_0\le LF_s$ at the start of stage $s$, where
$1\le s\le k$, after $b_{\rm start}$ edges have been read.
If the stage is executed, let its $j$th residual graph be obtained
by deleting the first $b_{\rm start}+(j-1)T_s$ edges. If all these
graphs have matching size at least $\nu_0/2$ and a strict majority
of the $r_s$ candidates satisfy
$c_*(T_s/m_0)\nu_0\le\lambda\le\nu_0$, then
\[
 1\le L_{\rm new}\le\nu_0\le B_{s-1}L_{\rm new}.
\]
The same conclusion holds unconditionally for a skipped stage
$L\ge\kappa T_s$, with $L_{\rm new}=L$.
\end{lemma}
\begin{proof}
The median and its maximum with the old rate preserve the upper bound
by $\nu_0$. The lower bound and the floor in $T_s$ give
\[
 \frac{\nu_0}{L_{\rm new}}
 \le\frac{m_0}{c_*T_s}
 \le\frac{2r_s\Lambda(F_s)}{c_*\epsilon2^{-s}}
 \le\frac{8R}{Dc_*\epsilon}B_{s-1}
 \le B_{s-1}.
\]
Here $F_s\le B_s$,~\eqref{wrap:growth} applies, and the final
inequality is one of the parameter constraints. A skipped stage
satisfies the same lower estimate by the deterministic shortcut after
\eqref{phase:improvement}.
\end{proof}

\begin{theorem}[Known length and a deterministic prefix bound]
\label{wrap:known-theorem}
Use the constants of Section~\ref{sec:parameters}. For every simple
graph $G_0=([n],E_0)$ with $n\ge0$ and $m_0=|E_0|\ge1$,
Algorithm~\ref{alg:known} on input $m_0$ finishes its active phase
reads within the first $\lceil\epsilon m_0\rceil$ edges on every
permutation and coin assignment. Its integer candidate $Z$ is
determined by those edges and the private coins. Under a uniform
edge permutation and independent private coins,
\begin{equation}
 \Prb\{\nu_0/B_0\le Z\le\nu_0\}
 \ge1-2\epsilon-2^{-6}>0.98,
 \label{wrap:known-success}
\end{equation}
where $\nu_0=\nu(G_0)$. Stages and copies use consecutive windows,
reset the sampler to idle at each phase horizon, and run sequentially;
only one recursion stack is live at any time.
\end{theorem}
\begin{proof}
The proof has two parts. A short prefix preserves a constant
fraction of one fixed maximum matching. Conditional phase guarantees
then control the first stage that fails to improve the estimate.
We use these events together without conditioning a phase on future
matching preservation.

Fix a maximum matching $M^*$ of size $\nu_0$. Its expected number
of edges in the first $\lfloor\epsilon m_0\rfloor$ positions is
at most $\epsilon\nu_0$. By Markov, with failure at most
$2\epsilon$, at most half these matching edges are read. On the
complementary event every residual graph in~\eqref{wrap:prefix}
has matching size at least $\nu_0/2$.
We do not condition a phase on that future preservation event.
Instead condition each phase on its complete actual past, including
all earlier edges and coins. This fixes its residual graph and its
parameters, while the unread suffix remains uniformly ordered.
Whenever the current hypotheses of~\eqref{phase:improvement} hold,
its conditional failure probability is at most $1/10$.
For copy $j$ of stage $s$, define an artificial failure $X_{s,j}$:
it is the indicator that~\eqref{phase:improvement} fails if the
previous-stage invariant and current residual-matching bound hold
at its start, and zero otherwise. Use zero indicators for skipped
copies. These analytical indicators are adapted and satisfy
$\E[X_{s,j}\mid\text{past}]\le1/10$. Iterating the conditional
exponential-moment bound at parameter $\log9$ gives
\begin{equation}
 \Prb\left\{\sum_{j=1}^{r_s}X_{s,j}\ge r_s/2\right\}
 \le(3/5)^{r_s}\le2^{-s-6}.
 \label{wrap:failure}
\end{equation}
No independence of the copies or stages is assumed.
On the preservation event, consider the first stage that does not
advance its invariant. All previous invariants then hold. By
Lemma~\ref{wrap:stage}, that stage must have a strict majority of
artificial failures. A union bound over this first bad stage,
together with the preservation failure, proves
\eqref{wrap:known-success}. Initially $L=1$ and $F_k=m_0$ are valid;
after stage one the factor is $B_0$. The branch $m_0\le B_0$
is correct deterministically. The prefix and space assertions follow
from~\eqref{wrap:prefix} and the sequential execution.
\end{proof}
\section{Unknown length, space, and running time}
\label{sec:unknown}

Theorem~\ref{wrap:known-theorem} gives more than an early stopping
bound: the candidate is fixed by a short prefix on \emph{every}
execution. This property permits a one-instance wrapper.
Set $\eta=\epsilon$ and choose a dyadic constant $\xi\ge\eta$.
Define prescribed sizes
\[
 s_1=1,\qquad s_{j+1}=s_j+\lceil\xi s_j\rceil\quad(j\ge1),
 \qquad h_s=\lceil\eta s\rceil.
\]
\begin{algorithm}[t]
\caption{$\proc{UnknownLength}$}
\label{alg:unknown}
\small
\begin{enumerate}[leftmargin=1.7em,label=\arabic*.,itemsep=3pt]
\item Index global stream positions by $t=1,2,\ldots$; boundary
$t$ is reached after edge $t$ is read. At boundary zero the saved
answer is $1$ and no instance is active.
\item At every boundary $t\ge1$, first commit any active instance whose
commitment is due: at boundary $s_j+h_{s_j}$, save its answer and
erase its workspace. Keep only this last committed answer.
\item If $t=s_j$ for some $j\ge1$, start
$\proc{KnownLength}(s_j)$ on global positions
$s_j+1,\ldots,s_j+h_{s_j}$ with fresh independent coins.
Each phase uses its own local clock starting at boundary zero.
The candidate becomes fixed within this interval. A commitment at this
same boundary is processed first, as in step 2.
\item At EOF, return zero if the stream is empty; otherwise return
the saved answer. Discard any uncommitted instance.
\end{enumerate}
\end{algorithm}
Because $h_s\le\lceil\xi s\rceil$, the half-open active
intervals $[s_j,s_j+h_{s_j})$, $j\ge1$, are pairwise disjoint.
They may share a boundary, where commitment precedes the next start.
Computing a candidate early never changes its commitment boundary.

The observed block of the last committed instance need not be an
initial block of the whole stream. Its positions are fixed by $m$,
however, so it is a uniform ordered sample without replacement.
Completing it to a larger uniform subgraph in the analysis permits
an application of the known-length theorem.

\begin{lemma}[Virtual completion]\label{wrap:virtual}
For the constants in Section~\ref{sec:parameters} and every simple
graph $G=([n],E)$, $n\ge0$, in uniformly random edge order,
the output $\widehat\nu$ of Algorithm~\ref{alg:unknown} satisfies
\[
 \Prb\{\nu(G)/2^{142}\le\widehat\nu\le\nu(G)\}>0.98.
\]
On every execution, at most one known-length instance is active,
and the wrapper uses $O(\log(n+2))$ additional bits.
\end{lemma}
\begin{proof}
Write $\nu=\nu(G)$. The cases $m=0,1$ are exact. For $m\ge2$,
let $s$ be the last
committed prescribed size. This size and its active interval depend
only on $m$, not on the input labels or previous outputs.
The tuple of $h_s$ observed edges is a uniform ordered tuple of
distinct edges of $G$. For analysis, append $s-h_s$ distinct edges
uniformly from its complement, independently of the instance's coins
conditional on that tuple. This gives a uniformly ordered $s$-edge
subgraph $H$ of $G$. The executed candidate is exactly the
known-length candidate on $H$, because that candidate uses only its
first $h_s$ edges. The algorithm makes no such completion.
We bound $p=s/m$ when $\nu(G)>B_0$. The next size is
$s'=s+\lceil\xi s\rceil$ and its commitment has not occurred.
Thus
\begin{equation}
 m<s'+\lceil\eta s'\rceil
 \le(1+\xi)(1+\eta)s+2+\eta.
 \label{wrap:next-size}
\end{equation}
The chosen constants and $m> B_0$ imply $s>B_0/2$ and then
$m/s<1+2\xi$. Hence
\[
 p>1-2\xi=1-2^{-41}.
\]
For a fixed maximum matching $M$ of $G$, let $X=|M\cap E(H)|$.
Its hypergeometric mean is $p\nu$ and its variance is at most
$p(1-p)\nu<2^{-41}\nu$. With $\tau=2^{-40}$, Chebyshev gives
\[
 \Prb\{X<(1-\tau)\nu\}
 <2^{41}/\nu<2^{-100}.
\]
Intersect this event's complement with the known-length success
event on $H$. Since $\nu(H)\le\nu(G)$, the saved answer satisfies
\[
 \frac{(1-\tau)\nu(G)}{B_0}\le Z\le\nu(G).
\]
The ratio is $B_0/(1-\tau)<2^{142}$, and the total failure is
at most $2\epsilon+2^{-6}+2^{-100}<0.02$.
For $1\le\nu(G)\le B_0$, the minimum answer one supplies the
lower bound and the known-length success event supplies the upper
bound. No union bound over the earlier committed instances is needed.
The sizes, global position, commitment boundary, and saved answer
are constantly many $O(\log(n+2))$-bit integers. Only one instance
is active because the intervals are disjoint.
\end{proof}
\subsection{Implementation bounds}

We show that the high-level algorithm has a uniform word-RAM
implementation with the same output law. Its only randomized
operations are dyadic coins and uniform choices from power-of-two
integer intervals, so they can be implemented exactly by fair bits.

\begin{proposition}[Worst-case implementation bounds]\label{prop:implementation}
With the constants of Section~\ref{sec:parameters},
Algorithm~\ref{alg:unknown} has a uniform one-pass implementation
using integer arithmetic and fresh fair random bits on
$\Theta(\log(n+2))$-bit words fixed from $n$ alone. For every
$n\ge0$, simple graph $G=([n],E)$, and permutation of its edges,
the implementation has exactly the algorithm's output distribution.
For every coin sequence, it halts using $O(\log^2(n+2))$ bits of
working memory and $O((m+1)\log(n+2))$ word operations, including
stream reads and random-bit generation. It outputs zero exactly
when the stream is empty and a positive integer otherwise. All
implicit constants are uniform over inputs and coin sequences.
\end{proposition}
\begin{proof}
The maximum recursion level is $O(\log(m+2))$, since every cap is
at most its phase length. Each frame contains constantly many
vertex names, integer clocks, and residuals of
$O(\log(n+2))$ bits. Simplicity gives $m\le\binom n2$, so the
stack uses $O(\log^2(n+2))$ bits on every execution.
The outer schedule needs only $O(\log(m_0+2))$ bits. Indeed,
the growth $B_s\ge B_{s-1}^2$ implies
\[
 \sum_{s=0}^{k-1}\log_2 B_s\le2\log_2B_{k-1}<2\log_2m_0.
\]
Self-delimiting field lengths add
$O(k\log\log(m_0+2))=O(\log(m_0+2))$ bits.
At a stage there are $O(k)$ dyadic candidate exponents, each of
$O(\log\log(m_0+2))$ bits, and their median can be computed in
place. All remaining clocks, exact output counts, and rates are
constantly many integers of $O(\log(m_0+2))$ bits.
No bound relies on the estimates being correct.
For running time, each physical edge causes at most
$O(\log(n+2))$ frame updates. The dispatcher can pop or cancel
at most the stack depth, and can then launch at most one new call:
every new call has a strictly later endpoint. Calendar alignment,
residual subtraction, guards, and dyadic comparisons use constant
word operations per frame. Drawing a threshold or birth coin uses
$O(\log(n+2))$ fresh fair bits; the blind branch uses a fixed
number. Thus the sampler takes $O(\log(n+2))$ time per processed
edge, including random-bit generation.
Target generation and median computations take a fixed polynomial
in $\log(m_0+2)$ word operations, for example by insertion-sorting
the candidate exponents. This is bounded by $O((m_0+1)\log(n+2))$.
In the unknown-length wrapper, completed active intervals have
length at least $\eta s$, so disjointness gives
$\sum s\le m/\eta$ over completed instances. A final uncommitted
instance has prescribed size at most $m$. Since $\eta$ is an
absolute constant, their total setup and processing time is
$O((m+1)\log(n+2))$. All operations use the integer clocks,
labels, residuals, and dyadic exponents already described; the
reference laws and proof-only monitor and tester are never evaluated.
Exact fair-bit sampling gives the same transition probabilities,
and hence the same output law, on every fixed input sequence.
The initialization and final output rule give the sign guarantee.
\end{proof}

\begin{proof}[Proof of Theorem~\ref{thm:main}]
The i.i.d. proxy and expected-charge estimates are
Propositions~\ref{iid:proxy} and~\ref{iid:fee}. The retained-state
comparison is proved in Section~\ref{sec:transfer}; applying it
with the bounded tester gives Proposition~\ref{phase:proposition}.
Theorem~\ref{wrap:known-theorem} supplies the known-length estimate
and its deterministic prefix bound. Lemma~\ref{wrap:virtual}
then gives the unknown-length estimate with $C=2^{142}$ and success
probability exceeding $0.98$, which implies the stated $2/3$.
The implementation bounds above establish the claimed worst-case
space and total running time. All constants and floor inequalities
used in these implications are verified next.
\end{proof}
\section{A sufficient parameter choice}
\label{sec:parameters}

We collect the constants here so that the algorithm and its analysis
can be read without optimizing a numerical approximation ratio.
The choices are deliberately conservative. They give the explicit
constant $C=2^{142}$ in Theorem~\ref{thm:main}; the result of interest
is that $C$ is independent of the graph size.
\begin{table}[ht]
\centering\small
\begin{tabular}{@{}p{.25\linewidth}p{.39\linewidth}p{.25\linewidth}@{}}
\toprule
Part & Parameters & Choice\\
\midrule
Recursive tests & $(c,\delta,\theta)$ & $(2^{12},2^{-5},2^{-165})$\\
Occupation & $B=1152(c+1)$, $\gamma=\delta/8$ & $B<2^{23}$, $\gamma=2^{-8}$\\
Moment and charge & $P=8B/\gamma+4$, $K_0$ & $P<2^{34}$, $K_0=2^{28}$\\
Phase cap and information & $(\kappa,B_{\rm mon},a)$ & $(2^{-15},2^{30},2^{-42})$\\
Count interval & $(\ell,u)$ & $(2^{-5},8)$\\
Tester & $(b_{\rm test},\zeta)$ & $(2^{11},2^{-7})$\\
Candidate normalization & $(S,c_*)$ & $(2^{33},2^{-79})$\\
Refinement & $(\epsilon,R,D,B_0)$ & $(2^{-44},16,2^{130},2^{141})$\\
Unknown length & $(\eta,\xi,\tau)$ & $(2^{-44},2^{-42},2^{-40})$\\
\bottomrule
\end{tabular}
\caption{All executable parameters are absolute dyadic constants.
The quantities $B,P,K_0$, the monitor, and the tester are used only in the proof.
}
\label{tab:parameters}
\end{table}
\paragraph{Recursive and survival constraints.}
We have $\delta\le1/24$ and $1/(c-1)<\delta$.
For $j\ge2$, $M_j=2^{12(j-1)-5}$, so exact threshold sampling
uses a fixed number of fair bits. For the second-moment proof,
$c\ge4$, $\delta/6=1/192\le\delta-1/40$, and
$3200<c-1=4095$. These are precisely the numerical conditions in
the low-load argument.
For the high-load envelope,
$(c+1)^4<2^{49}$ and $e^{-36}+\theta<2^{-50}$.
The occupation norm satisfies $\sqrt{6\delta}=\sqrt{3}/4<1/2$.
Finally, $B<2^{23}$ and $P<2^{34}$ imply the total root charge
bound $K_0=2^{28}$ by the explicit calculation in
Proposition~\ref{iid:fee}.
\paragraph{Phase constraints.}
The i.i.d. count error is below $32\kappa=2^{-10}$.
The tester cap is $C_{\rm test}=2^{16}+1$, and
$\log(C_{\rm test}+1)<64$. Its stated Chernoff errors are below
$\zeta=2^{-7}$. The monitor and information constraints are
\[
 \frac{K_0}{B_{\rm mon}}(1/7+\kappa)<3/64,
 \qquad 4B_{\rm mon}a=2^{-10}.
\]
Also $S=2^{33}>2^7(1+2^{25})$, and
\[
 \frac{\delta\kappa}{128c}=2^{-39},\qquad
 2^{-5}\cdot2^{-39}\cdot\frac12\cdot\frac1{2S}=c_*.
\]
For $T\ge2^{32}$, $\kappa T\ge2^{17}>H_1$.
These verify every numerical inequality in
Proposition~\ref{phase:proposition}.
\paragraph{Target growth.}

To prove~\eqref{wrap:growth}, it suffices that
\begin{equation}
 \frac{B_{s-1}}{D(s+1)2^s}
 \ge2\log_2B_{s-1}+1.\label{param:growth}
\end{equation}
At $s=1$, the two sides are $512$ and $283$.
The function $x/(2\log_2x+1)$ is increasing for $x\ge4$.
At the comparison points $x_s=2^{141\cdot2^{s-1}}$, the
denominator $D(s+1)2^s(2\log_2x_s+1)$ grows by a factor
less than six from one index to the next, while
$x_{s+1}/x_s=x_s>6$. Thus~\eqref{param:growth} propagates
at these points, and induction gives it at the actual targets.
Taking the floor still gives $e_s\ge2\log_2B_{s-1}$,
so $B_s\ge B_{s-1}^2$. Since the argument of the floor exceeds
one, $\Lambda(B_s)=1+e_s\le2B_{s-1}/[D(s+1)2^s]$.
\paragraph{Horizon floors and refinement.}

Put $q_0=\log_2m_0>141$. For $s\le k$,
$r_s\le32(s+1)\le32\cdot2^s$ and
$\Lambda(F_s)\le q_0+2\le2q_0$.
Using~\eqref{wrap:k-bound}, the quantity before the floor in $T_s$
is greater than $\epsilon m_0/q_0^3$.
The function $2^q/q^3$ is increasing for $q\ge141$, and
$141^3<2^{22}$, so this quantity exceeds
$2^{-44}2^{141}/2^{22}=2^{75}$.
Every $T_s\ge2^{75}$, and every floor loss used earlier is
therefore at most a factor of two.
The information condition holds because
$\epsilon2^{-s}/[r_s(1-\epsilon)]<2\epsilon<a$.
The invariant advances because
\[
 D=\frac{8R}{c_*\epsilon}.
\]
For the median error,
$(3/5)^{32(s+1)-1}\le2^{1-16(s+1)}\le2^{-s-6}$.
Summing over $s\ge1$ gives the claimed $2^{-6}$ failure budget.
\paragraph{Unknown length.}

We have $\eta=\xi/4$, so $h_{s_j}\le\lceil\xi s_j\rceil$.
The half-open intervals $[s_j,s_j+h_{s_j})$ are therefore pairwise
disjoint for all $j\ge1$.
If $\nu(G)>B_0$,~\eqref{wrap:next-size} first implies $s>B_0/2$:
otherwise its right side is less than $3B_0/4+3<B_0$.
Furthermore
\[
 \xi+\eta+\xi\eta<3\xi/2,
 \qquad (2+\eta)/s<(4+2\eta)/B_0<\xi/2.
\]
Thus $m/s<1+2\xi$ as used in Lemma~\ref{wrap:virtual}.
The total failure is at most
$2^{-43}+2^{-6}+2^{-100}<0.02$, and
$B_0/(1-\tau)<2^{142}$.
\paragraph{Space beyond the theorem.}
The second logarithmic factor in our implementation comes from the
simultaneous vertex identities and residuals in the recursive stack.
The scheduling and unknown-length reductions use only logarithmic
additional space. Whether the matching size admits a constant-factor
estimate in $O(\log n)$ bits from one random permutation remains open.
\clearpage
\appendix
\section{Sharpening the approximation constant}
\label{app:sharp}

The main proof leaves substantial slack so that each part can be read
independently. We now remove most of this slack without changing the
algorithmic framework. The vertex tests use smaller geometric scales;
the last reward increment is shortened to fit the phase; and each root
is padded for only three times its reward. Weighted load potentials
replace the polynomial survival envelope. Finally, binary relative
entropy gives more accurate phase failure probabilities than Pinsker's
inequality. This appendix proves each modified estimate, including the
changes to the reference distribution and its charges.
\begin{theorem}[Explicit constant]\label{sharp:thm:main}
Theorem~\ref{thm:main} holds with $C=2^{28}$.
\end{theorem}

The first two subsections establish the sharper sampler bounds.
The last subsection specifies the phase parameters and the full outer
schedule, verifies their inequalities, and proves the theorem. Throughout
this appendix the fixed blind-PASS probability is $\theta=2^{-200}$.
All other definitions refer to the modified parameters stated here.
\subsection{The sharpened i.i.d. proxy and information charges}
\label{sharp:core}

We use the same aligned vertex test with
\[
 c=512,\qquad \delta=\frac1{16},\qquad
 0<\theta\le\theta_*:=2^{-200},\qquad
 \bar c=\frac{511}{512},\qquad \rho=\frac{319}{512}.
\]
The fixed choice $\theta=2^{-200}$ suffices for the executable algorithm.
The canonical-call and child-budget proofs are unchanged: the scales
still divide one another, $M_j=\delta K_j$ is a power of two for
$j\ge2$, and $H_{j-2}<\delta K_j<H_{j-1}$.
The high-sample estimate is adjusted below rather than invoking the
convenient restriction $\delta\le1/24$ used earlier.
A root now receives an integer cap $h'\ge c+1$.  Write
$h'=H_J+r$, where $J\ge1$ is maximal and $0\le r<c^{J+1}$.
Test both endpoints through level $J+1$, stopping at the first failure.
If $I$ is the last level passed by both, its reward is $H'_I=H_I$
for $I\le J$ and $H'_{J+1}=h'$.  Thus its reward increments are
\[
 \iota_i=c^i\quad(0\le i\le J),\qquad \iota_{J+1}=r.
\]
Retain the definitions of $q_i,p_i,D_i,a_i$ from Section~\ref{sec:iid},
now for the new parameters, and put
\[
 z_i(uv)=p_i(u)p_i(v),\qquad
 \mu'=\frac1m\sum_{uv\in E}\sum_{i=0}^{J+1}\iota_i z_i(uv).
\]
Canonical sampling identifies $\mu'$ with the expected root reward.
We will prove the upper proxy bound $\mu'\le635\nu(G)$ and the
expected-charge bound $300\mu'$ used in the sharpened phase analysis.
\subsubsection{One survival envelope}

We first derive an exponential survival bound with explicit
constants. It improves the polynomial envelope used in the main proof
and will make the weighted-load potentials substantially smaller.
On an ordinary PASS, the direct-child budget leaves at least
$(1-6\delta)K_j-1\ge\rho K_j$ completed comparison marks when
$j\ge2$.  Each mark is at most $1/\bar c$, and their total mean is
at least $\rho a_{j-1}(v)$.
The normalized threshold $X=U_j/K_j$ lies in $[\delta,2\delta)$,
has mean at most $3\delta/2$, and is independent of the i.i.d. comparison
sequence.  For $s>0$, the chord bound and the power-series bound on
$\cosh$ give
\[
 \E[e^{sX}]
 \le\tfrac12(e^{s\delta}+e^{2s\delta})
 \le\exp(3s\delta/2+s^2\delta^2/8).
\]
For a mark $Y\in[0,1/\bar c]$,
$\E[e^{-\bar c tY}]\le\exp[-\bar c(1-e^{-t})\E[Y]]$.
Take $t=5/2$ and $s=\bar c t$.  Since
\[
 3s\delta/2+s^2\delta^2/8\le1945/8192<6/25,
 \qquad \bar c\rho(1-e^{-5/2})>
              \bar c\rho(11/12)>57/100,
\]
exponential Markov proves
\begin{equation}\label{sharp:eq:envelope}
 q_j(v)\le\theta+\exp\bigl(6/25-57a_{j-1}(v)/100\bigr).
\end{equation}
At level one, $q_1(v)\le\theta+e^{-a_0(v)}$, which also satisfies
\eqref{sharp:eq:envelope}.
For the stronger low-load estimate below, we use the following
oversampling inequality of KMNT~\cite[Lemma 1]{KMNT20}. We include
its short proof to keep the dependence on the parameters explicit.

The inequality says that averaging $c$ independent copies sharply
reduces a threshold-crossing probability when the mean is well below
that threshold. We use it to sum failure probabilities over levels.

\begin{lemma}[Oversampling]\label{iid:oversampling}
Let $X$ be a finite sum of independent random variables in $[0,1]$,
with $\E[X]\le d/3$.  Let $X_1,\ldots,X_c$ be independent copies.
If $t=cd/10\ge\max\{2,d+2\}$, then
\[
 \Prb\left(c^{-1}\sum_{i=1}^cX_i\ge d\right)
 \le\left(ce^{-3cd/10}+\frac{200}{9c}\right)\Prb(X\ge d).
\]
\end{lemma}
\begin{proof}
Write $X=\sum_a Z_a$ and $p=\Prb(X\ge d)$.  For $s>0$,
independence and $0\le Z_a\le1$ give
\[
 s\Prb(X\ge s)
 \le\sum_a\E\left[Z_a\Prb(X-Z_a\ge s-Z_a\mid Z_a)\right]
 \le(\E[X])\Prb(X\ge s-1).
\]
Applying this at $s=t,t-1$ yields
$\Prb(X\ge t)\le200p/(9c^2)$, since $t-2\ge d$ and
$t(t-1)\ge t^2/2$.
If $\sum_iX_i\ge cd$, either some $X_i>t$, or some
$X_i\in[d,t]$ while $\sum_{j\ne i}X_j\ge9cd/10$.
The latter sum is a sum of independent $[0,1]$ variables of mean
at most $cd/3$.  Exponential Markov at parameter one bounds its
upper tail by
$\exp[-(9/10-(e-1)/3)cd]\le e^{-3cd/10}$.
Independence from $X_i$ and a union bound finish the proof.  No division
by $p$ is used, so the assertion also holds when $p=0$.
\end{proof}
For low loads, apply Lemma~\ref{iid:oversampling} to
$\bar cS_{i-1}$ with $d=\bar c\delta$.
Its hypotheses hold when $a_{i-1}\le\delta/3$, since
$cd/10=511/160>d+2$.  The relative-tail coefficient obeys
\[
 \varpi=512e^{-1533/160}+\frac{200}{4608}
 <\frac{512}{12000}+\frac{200}{4608}<\frac1{11}.
\]
In the decomposition $S_i=A_i+B_i$, the top contribution $B_i$ is
integer-valued and $A_i$ is the average of $c$ copies of $S_{i-1}$.
Writing $\alpha_i=\Prb(A_i\ge\delta)$ and
$\beta_i=\Prb(B_i\ge1)$ gives
$\alpha_i\le\varpi(\alpha_{i-1}+\beta_{i-1})$ and
$\sum_{i\le r}\beta_i\le a_r$.
The low sample sandwich and summation therefore yield
\begin{equation}\label{sharp:eq:low-load}
 1-p_{r+1}(v)\le\kappa_{\mathrm{low}} a_r(v)
 \quad\text{if }a_r(v)\le1/48,\qquad \kappa_{\mathrm{low}}=11/10.
\end{equation}
The base-level incidence term is included in this sum.
\subsubsection{A weighted-load potential}

To bound a vertex's total weighted degree, we consider abstract
load/survival sequences satisfying the estimates just proved. This
separates the numerical potential calculation from the graph argument.
Let $C=c+1=513$ and
$\psi(x)=\min\{1,\theta_*+e^{6/25-57x/100}\}$.
We call a finite sequence $(x_i,p_i)$ a profile if $p_0=1$,
$0\le x_0\le1$, and, for $i\ge1$,
\[
 x_{i-1}\le x_i\le Cx_{i-1},\qquad
 0<p_i\le p_{i-1}\psi(x_{i-1}),\qquad
 p_i\ge1-\kappa_{\mathrm{low}} x_{i-1}\quad\text{when }x_{i-1}\le1/48.
\]
Set $x_{-1}=0$ and $\Delta x_i=x_i-x_{i-1}$.
The sequences $(a_i(v),p_i(v))$ are profiles by
\eqref{sharp:eq:envelope}--\eqref{sharp:eq:low-load}.
Shortening the last load increment also preserves all three conditions.

The potential $\Phi$ below pays for gradual load increases.
Above a fixed cutoff $H$, survival contracts fast enough to sum
the entire remaining contribution geometrically. The lemma combines
these two ranges.

\begin{lemma}[Weighted-load potential]\label{sharp:lem:potential}
Suppose $x_{i-1}\le x_i\le Cx_{i-1}$ for $i\ge1$, $0\le x_0\le1$, and
$r_0=1$, $0\le r_i\le r_{i-1}q(x_{i-1})$, where $q$ is
nonincreasing and takes values in $[0,1]$.
Let $F\ge1$ be increasing with $F(0)=1$, and put
$\Phi(x)=\int_0^xF(z)\,dz$.
Suppose $x'F(x')\le D xF(x)$ for $x\le x'\le Cx$.
Choose $H>1$ with $Dq(H)\le10^{-3}$ and put $k=1000/999$.
If
\[
 M\ge\Phi(H),\qquad
 M\ge\sup_{0\le x<H}\{\Phi(x)+cxF(x)q(x)\}
           +kc\sup_{H\le x\le CH}xF(x)q(x),
\]
then $\sum_i r_iF(x_{i-1})\Delta x_i\le M$.
\end{lemma}
\begin{proof}
For the maximum remaining cost over $n$ steps from load $x$ and
current weight one,
\[
 V_n(x)\le q(x)\sup_{x\le x'\le Cx}
              [(x'-x)F(x)+V_{n-1}(x')],\qquad V_0=0.
\]
For $x\ge H$, induction gives $V_n(x)\le kcxF(x)q(x)$,
using $x'-x\le cx$, the growth bound on $xF(x)$, and
$1+kDq(x)\le1+k/1000=k$.
For $x<H$, inductively prove $\Phi(x)+V_n(x)\le M$.
If $x'<H$, use
$(x'-x)F(x)\le\Phi(x')-\Phi(x)$.
If $x'\ge H$, then $x'\le CH$, and the stated supremum bound
and the high-load estimate apply.  Finally, the initial contribution
$x_0$ is at most $\Phi(x_0)$.
\end{proof}
Apply this lemma with $r_i=p_i$, $q=\psi$, $F=1$, $D=C$,
and $H=24$.
The inequality $e^{336/25}>600000$ gives
$C\psi(H)<10^{-3}$.
The high-region term and the positive-floor part of the low region
are together at most
\[
 R=kc(24/600000+24C\theta_*)+24c\theta_*<1/40.
\]
The function $xe^{6/25-57x/100}$ is maximized at $x=100/57$ and
decreases afterward.  Therefore the low-region supremum plus $R$
is bounded, in its three respective ranges, by
\[
 \begin{array}{c|c}
 x\in[0,2]&2+\dfrac{51200}{57}e^{-19/25}+\dfrac1{40}
                <2+\dfrac{8000}{19}+\dfrac1{40}\\[5pt]
 x\in[2,4]&4+1024e^{-9/10}+\dfrac1{40}
                <4+\dfrac{20480}{49}+\dfrac1{40}\\[5pt]
 x\in[4,24)&24+2048e^{-51/25}+\dfrac1{40}
                <24+\dfrac{4096}{15}+\dfrac1{40}.
 \end{array}
\]
All three upper bounds are below $1270/3$.  Hence every profile satisfies
\begin{equation}\label{sharp:eq:load}
 \sum_i p_i\Delta x_i\le1270/3.
\end{equation}

The preceding weighted-degree bound yields a better upper proxy
through fractional matching. For the lower proxy, a vertex whose
survival falls below $1-\tau$ must already have accumulated load
$\tau/\kappa_{\mathrm{low}}$. Counting these vertices gives
a coefficient close to one when $h'/m$ is small.

\begin{proposition}[Sharpened matching proxy]\label{sharp:prop:proxy}
The truncated reward satisfies $\mu'\le635\nu(G)$.
For every $0<\tau\le\kappa_{\mathrm{low}}/48$,
\begin{equation}\label{sharp:eq:proxy-lower}
 \mu'\ge
 \frac{(1-\tau)^2}{1+2(1-\tau)\kappa_{\mathrm{low}} h'/(\tau m)}
          \frac{h'}m\nu(G).
\end{equation}
In particular, if $h'/m\le1/49152$, then
$\mu'\ge(49/50)(h'/m)\nu(G)$.
\end{proposition}
\begin{proof}
Give edge $uv$ weight $w_{uv}=m^{-1}\sum_i\iota_i z_i(uv)$.
Its weighted degree at $v$ is
$\sum_i p_i(v)\Delta a'_i(v)$, where only the last increment is
shortened, by the factor $r/c^{J+1}$.  This remains a profile, so
\eqref{sharp:eq:load} bounds every weighted degree by $1270/3$.
Scaling by this number gives a fractional matching with total weight
$\mu'/(1270/3)$.
The degree-constraint matching polytope has integrality gap at most
$3/2$: its extreme points consist of unit edges and odd cycles of
half-weight edges, and rounding each cycle loses at most a factor
$3/2$.  For completeness, after removing unit edges, any positive-support
tree has at least two slack leaves, and any component with more than
one cycle has more variables than degree equations.  A remaining
unicyclic component must be a cycle with all degree equations tight;
an even cycle admits an alternating perturbation, and an odd cycle
has all edge values $1/2$.  Thus $\mu'\le635\nu(G)$.
For the lower bound let $\gamma=\tau/\kappa_{\mathrm{low}}$ and
$C'=\{v:p_{J+1}(v)\le1-\tau\}$.
Every $v\in C'$ has $a_J(v)\ge\gamma$, by
\eqref{sharp:eq:low-load}.  At its first crossing of $\gamma$,
its survival is at least $1-\tau$, so its weighted load through
level $J$ is at least $(1-\tau)\gamma$.
Consequently $|C'|\le2\mu'/[(1-\tau)\gamma]$.
At least $\nu(G)-|C'|$ edges of a maximum matching avoid $C'$,
and each contributes at least $(1-\tau)^2h'/m$ to the reward.
Substitution and rearrangement prove \eqref{sharp:eq:proxy-lower}.
For the final assertion take $\tau=1/256$; the displayed coefficient
is at least $65025/66284>49/50$.
\end{proof}
\subsubsection{Three logarithmic moments}

The refined state comparison uses both survival logarithms and
logarithms of local proposal normalizers. We bound their first moments
and one higher moment; the latter controls recursive descendants by
H\"older's inequality. Put $\zeta=201/100$. For a profile let
\[
 S=\sum_i p_i\Delta x_i,\qquad
 L_i=\log(1/p_i),\qquad
 Z_i=L_i+\log(1+\zeta x_{i-1}).
\]

\begin{lemma}[Profile moments]\label{sharp:lem:moments}
Every profile satisfies
\begin{equation}\label{sharp:eq:moments}
 \sum_i p_iL_i\Delta x_i\le16S,\qquad
 \sum_i p_iZ_i\Delta x_i\le20S,\qquad
 \sum_i p_iZ_i^8\Delta x_i\le25^8S.
\end{equation}
\end{lemma}
\begin{proof}
Let $\eta=2^{-50}$ and $q=\psi^{1/4}$.  The positive blind-PASS
floor is retained in the bound
\[
 q(x)\le\eta+e^{3/32-x/8}\le\eta+e^{1/8-x/8}.
\]
Apply Lemma~\ref{sharp:lem:potential} to $r_i=p_i^{1/4}$ in the
following two cases:
\[
\begin{array}{c|c|c|c|c}
 F(x)&H&D&\text{upper bound on }\Phi(H)&M\\\hline
 1&112&C&112&2000\\
 (1+3x)^{3/4}&160&C^{7/4}&10000&36000.
\end{array}
\]
Here $Cq(112)<10^{-3}$ and $C^{7/4}q(160)<10^{-3}$, using
$e^{111/8}>10^6$, $C^{7/4}<56000$, and
$e^{159/8}>4\cdot10^8$.
For $F=1$, the exponential part of the low-region supremum, multiplied
by $c$, is at most $4096e^{-7/8}<1800$.
For the second function, put $y=x+1/3$ and maximize
$y^{7/4}e^{-y/8}$ at $y=14$ to obtain
\[
 c\sup_{x>0}x(1+3x)^{3/4}e^{1/8-x/8}
 \le512\cdot3^{3/4}14^{7/4}e^{-19/12}<25000.
\]
In either case the positive-floor contribution to the low region is
below one.  The complete high-region contributions are respectively
at most
\[
 kc(112/10^6+112C\eta)<1,
 \qquad
 kc\left(\frac{160\cdot104}{4\cdot10^8}
                  +160C\cdot12000\eta\right)<1.
\]
For the latter inequality use $F(160)<104$, $F(160C)<12000$,
and the decrease of $xF(x)e^{-x/8}$ for $x\ge160$.
The potential lemma thus gives
\begin{equation}\label{sharp:eq:quarter-moments}
 \sum_i p_i^{1/4}\Delta x_i\le2000,\qquad
 \sum_i p_i^{1/4}(1+3x_{i-1})^{3/4}\Delta x_i\le36000.
\end{equation}
If the profile crosses $1/48$, survival through the first crossing is
at least $1-\kappa_{\mathrm{low}}/48>97/100$, so $S\ge97/4800$.
Give index $i$ probability $p_i\Delta x_i/S$ and denote the resulting
random variables by $L,Z$.
Since $\zeta<3$, \eqref{sharp:eq:quarter-moments} implies
\[
 \E[e^{3L/4}]\le\frac{2000\cdot4800}{97}<2^{17},\qquad
 \E[e^{3Z/4}]\le\frac{36000\cdot4800}{97}<2^{21}.
\]
If the profile never crosses $1/48$, then
$L\le\log(100/97)$ and
$Z\le\log[(100/97)(17/16)]$, so the same bounds hold.
The case $S=0$ is immediate.
Jensen's inequality and $\log2<7/10$ now give $\E[L]<16$ and
$\E[Z]<20$.
Writing $a=28\log2$, the second exponential-moment bound gives
$\Prb(Z>z)\le\min\{1,e^{-3(z-a)/4}\}$.
Thus $Z$ is stochastically dominated by $a+E$, with $E$ exponential
of rate $3/4$.  Minkowski's inequality and $8!<4^8$ yield
\[
 (\E[Z^8])^{1/8}
 \le28\log2+\tfrac43(8!)^{1/8}
 <\tfrac{98}{5}+\tfrac{16}{3}<25.
\]
This proves all three claims.
\end{proof}
\subsubsection{Occupation and the charge constant}

We combine the profile moments with a sharper bound on the
expected budget at each recursion depth. The root endpoint calls
and padding are treated separately, since their joint prerequisites
have smaller charges than a general recursive frame.

For types $(i,v)$, now with $1\le i\le J+1$, put
\[
 \omega_i(v)=\frac{K_i}{m}p_{i-1}(v)D_{i-1}(v),\qquad
 W=\frac1m\sum_e\sum_{a=0}^Jc^az_a(e)\le\mu'.
\]
Then $\sum\omega=2W$, and $\omega$ dominates initial root-call
budget.  The same offspring matrix satisfies
\[
 \sum_xT_{xy}\le3\delta,\qquad T\omega\le\rho^{-1}\omega.
\]
For the second bound, with $r=D_{j-1}(u)/m$, the ordinary call has
$t_j^{\rm ord}(u)r\le1-(1-r)^{K_j}$, whereas its PASS event completes
at least $\rho K_j$ comparison tuples without a highest-level survivor.
Hence
$1-q_j^{\rm ord}(u)\ge1-(1-r)^{\rho K_j}$ and
$t_j(u)r\le\rho^{-1}(1-q_j(u))$.
The level-one case is direct.  The telescoping argument from
Lemma~\ref{iid:occupation} proves the claimed row bound.
For $A_{xy}=T_{xy}\omega_y/\omega_x$, weighted Jensen gives, for
$s\ge1$,
\[
 \|A\|_{L^s(\omega)}\le
        \rho^{-(1-1/s)}(3\delta)^{1/s}.
\]
In particular,
\[
 \|A\|_{L^1(\omega)}\le3\delta,\qquad
 \|A\|_{L^{8/7}(\omega)}\le
 \alpha:=\rho^{-1/8}(3\delta)^{7/8}<1/4;
\]
the last inequality is equivalent to $2187<2552$.
As before, $g=(I-T)^{-1}\omega$ dominates actual occupation.
For
\[
 \psi_i(v)=\log(1/p_{i-1}(v)),\qquad
 \lambda_i(v)=\log(1+\zeta a_{i-2}(v)),\qquad u_i=\psi_i+\lambda_i,
\]
Lemma~\ref{sharp:lem:moments} gives
$\sum\omega\psi\le32W$, $\sum\omega u\le40W$, and
$\sum\omega u^8\le2\cdot25^8W$.
The $L^1$ bound and H\"older with exponents $8/7,8$ imply
\begin{equation}\label{sharp:eq:occupation-charge}
 \sum_xg_x\le\frac{2W}{1-3\delta},\qquad
 \sum_xg_xu_x\le2W\left(20+\frac{25\alpha}{1-\alpha}\right).
\end{equation}
Indeed, write $g/\omega=\sum_{k\ge0}A^k\mathbf1$, use the first
moment for $k=0$, and sum the geometric bound for $k\ge1$.
The modified charges, whose pathwise coverage is proved in
Section~\ref{sharp:transfer}, are as follows.
A non-root call pays $2K_i(c_0+\psi_i+\lambda_i)$, where
$c_0=1+\log12$.  For a root endpoint, replace its own $\psi_i$ by
the logarithm of the reciprocal joint entry probability.
Padding pays $3H'_I[1+\log(1/z_I(e))]$.  No birth charge is needed.

The next proposition collects these estimates for the exact charges
used by the sharpened reference construction. Its constant is uniform
in the graph and cap.

\begin{proposition}[Sharpened expected charge]\label{sharp:prop:fee}
Conditional on every i.i.d. pre-birth history, these total root charges have
expected total at most $300\mu'$.
\end{proposition}
\begin{proof}
By \eqref{sharp:eq:occupation-charge}, base call charges contribute at most
\[
 4W\left(\frac{c_0}{1-3\delta}+20+
                              \frac{25\alpha}{1-\alpha}\right).
\]
For a root edge $uv$ with $u$ first, put
$z=p_{i-1}(u)p_{i-1}(v)$ and $q=q_i(u)$.
The extra prerequisite logarithms, after weighting by entry
probabilities, are
$z\psi_i(v)+zq[\psi_i(u)+\log(1/q)]$.
Using $q\log(1/q)\le1/e$, their total surcharge is at most
$2(32+1/e)W$.
To bound padding, condition on a root edge and let
$\Lambda_i=\log(1/z_i)$ and $\pi_i=\Prb(I=i\mid e)$.
Thus $\pi_i=z_i-z_{i+1}$ for $i<J+1$, and
$\pi_{J+1}=z_{J+1}$.  Summation by parts gives
\[
 \sum_{i\ge a}\pi_i\Lambda_i
 =z_a\Lambda_a+\sum_{i>a}z_i(\Lambda_i-\Lambda_{i-1})
 \le z_a(\Lambda_a+1),
\]
since $z_i\log(z_{i-1}/z_i)\le z_{i-1}-z_i$.
Consequently
\[
 \E[H'_I\Lambda_I\mid e]
 \le\sum_a\iota_a z_a(1+\Lambda_a).
\]
After averaging over root edges, its logarithmic part is the first
moment of the shortened profiles.  Lemma~\ref{sharp:lem:moments}
bounds it by $32\mu'$.  Thus the complete PAD charge is at most
$3(2+32)\mu'=102\mu'$.
Finally, $W\le\mu'$, $c_0<7/2$, $\alpha<1/4$, and $1/e<3/8$
give a total coefficient below
\[
 2\left(\frac{112}{13}+40+\frac{50}{3}+32+\frac38\right)+102
 =\frac{46381}{156}<300.
\]
All bounds are uniform over $0<\theta\le\theta_*$.
\end{proof}
\subsubsection{Finite arithmetic for the analytic constants}

The exponential comparisons used above require no numerical
optimization.  Each row below satisfies
$\sum_{j=0}^{N}x^j/j!>b$, and hence proves $e^x>b$ by a rational
calculation.
\begin{center}
\begin{tabular}{@{}c r r@{}}
\toprule
$x$&$b$&$N$\\\midrule
$5/2$&$12$&$6$\\
$7/10$&$2$&$3$\\
$1$&$8/3$&$4$\\
$1533/160$&$12000$&$12$\\
$336/25$&$600000$&$18$\\
$19/25$&$32/15$&$4$\\
$9/10$&$49/20$&$4$\\
$51/25$&$15/2$&$5$\\
$111/8$&$10^6$&$20$\\
$159/8$&$4\cdot10^8$&$27$\\
$7/8$&$512/225$&$3$\\
$19/12$&$97/20$&$6$\\\bottomrule
\end{tabular}
\end{center}
For the second potential, the remaining comparisons follow by raising
to fourth powers:
\[
 513^7<56000^4,\quad481^7<52501^4,\quad
 481^3<104^4,\quad246241^3<12000^4.
\]
Also $3^{3/4}<57/25$ and $14^{7/4}<102$, so its low-region bound
follows from
$512(57/25)\,102(20/97)<25000$.
All other displayed comparisons are rational inequalities.
\subsection{The transfer bound with the sharpened constants}
\label{sharp:transfer}

The anchored-state argument also applies to the sharpened sampler.
We give the changes explicitly, including the reference normalizers and
the charges that replace the more generous constants in
Section~\ref{sec:transfer}.
Throughout this subsection,
\[
 c=512,\qquad \delta=1/16,\qquad 0<\theta\le2^{-200},
 \qquad T/m\le1/512.
\]
Put \(K_j=c^{j-1}\), \(H_i=\sum_{a=0}^ic^a\), and \(H_{-1}=0\).
For an integer cap \(h'\ge c+1\), let \(J\) be maximal with
\(H_J\le h'\), and put \(r=h'-H_J\).
The root tests its endpoints through level \(J+1\), stopping at the
first failure, and its reward is
\[
 H'_I=H_I\quad(0\le I\le J),\qquad H'_{J+1}=h'.
\]
For \(e=\{u,v\}\), write \(z_i(e)=p_i(u)p_i(v)\).  Thus
\[
 \mu'=\frac1m\sum_e\left(\sum_{i=0}^Jc^iz_i(e)+rz_{J+1}(e)\right),
 \qquad W_J\le\mu'.
\]
\paragraph{Duration and state.}

If the root is born at boundary \(b\), its padding deadline is
\(b+3H'_I\).  This is strictly after its natural decision.
Indeed, on the all-\(\PASS\) calendar, let \(E_j\) be the boundary
at which both level-\(j\) endpoint tests return.  The second test
has no alignment wait.  Since \(E_{j-1}\) is a multiple of
\(K_{j-1}\), the first level-\(j\) wait is at most \(K_j-K_{j-1}\).
Consequently
\[
 E_1=b+2,\qquad
 E_j-E_{j-1}\le3K_j-K_{j-1},\qquad
 E_j\le b+3H_{j-1}-1.
\]
Failure at level \(I+1\) occurs no later than \(E_{I+1}\);
success through the final level occurs at \(E_{J+1}\).
Both precede the stated deadline, including when \(r=0\).
The exact duration, including the birth edge, is therefore \(1+3H'_I\).
The cap and its truncated increment are fixed phase parameters.
The root's current level and endpoint role still determine its query
calendar.  In padding, its birth, edge, and terminal level determine
the deadline; no natural decision time or failure cause is retained.
Thus the state reconstruction and common-kernel argument are unchanged.
\paragraph{Sharper birth windows.}

Use the call and child proposals of Section~\ref{sec:transfer}, extending
the root levels through \(J+1\).  For a current level-\(j\) root
endpoint, all preceding levels consume at most \(4H_{j-2}\) positions.
The two level-\(j\) endpoint tests together consume at most \(3K_j\).
For either endpoint role there are therefore at most
\(3K_j+4H_{j-2}\) possible integer root births at a fixed observation
boundary.  Both endpoint prerequisites are at most \(z_{j-1}(e)\).
The total call-root proposal weight is at most
\begin{align}
 \frac{\lambda}{m}
 \sum_e\sum_{j=1}^{J+1}
       (6K_j+8H_{j-2})z_{j-1}(e)
 &\le
 \left(6+\frac8{c-1}\right)\lambda W_J
 \notag\\
 &\le
 \left(6+\frac8{c-1}\right)\lambda\mu'.
 \label{sharp:eq:call-mass}
\end{align}
The bound counts birth \(b=t\), when the first call has just been
created, and uses \(H_{j-2}<K_j/(c-1)\).
A classification currently at level \(i\) has elapsed time less than
\(2H_{i-1}\).  Its source birth therefore has at most \(2H_{i-1}\)
possible integer values.  This includes a child still in WAIT whose
return lies beyond an ancestor deadline; only the observation boundary
must precede that ancestor deadline.
At a parent of type \((j,v)\), with \(k\) previously chosen anchors,
the local child normalizer consequently satisfies
\begin{align}
 \mathcal N-1
 &\le\frac{2c}{(c-1)(m-k)}
             \sum_{i<j}K_iD_{i-1}(v) \notag\\
 &\le\frac{2c}{c-1}\frac{m}{m-T}\,a_{j-2}(v)
 \le \zeta\,a_{j-2}(v),\qquad \zeta=\frac{201}{100}.
 \label{sharp:eq:child-normalizer}
\end{align}
For \(j=1\) the sum is empty and \(a_{-1}=0\).
The last comparison is exact:
\[
 \frac{2c}{c-1}\frac{m}{m-T}
 \le \frac{524288}{261121}<\frac{201}{100}.
\]
The padding option \((\PAD,b,e,i)\) has weight
\(\lambda z_i(e)/m\), with
\(b\le t<b+3H'_i\).  It has at most \(3H'_i\) possible births.
For each fixed edge, let
\(M_e=\sum_{i=0}^Jc^iz_i(e)+rz_{J+1}(e)\).
Then
\[
 \sum_{i=0}^{J+1}H'_iz_i(e)
 \le \frac{c}{c-1}\sum_{i=0}^Jc^iz_i(e)
      +H_Jz_{J+1}(e)+rz_{J+1}(e)
 \le \frac{2c}{c-1}M_e.
\]
Here \(z_{J+1}\le z_J\) bounds the \(H_Jz_{J+1}\) term,
and \(c/(c-1)>1\) covers the truncated increment.
Thus the total padding weight is at most
\(6c\lambda\mu'/(c-1)\).
With idle weight one and dead weight one, the root normalizer is
\begin{equation}\label{sharp:eq:root-normalizer}
 Z_0(t)\le2+\zeta_0\mu'/L,\qquad
 \zeta_0=6+\frac8{511}+\frac{3072}{511}<12.03.
\end{equation}
The possible overlap of padding windows at different terminal levels,
including \(r=0\), causes no difficulty: these are separate
alternatives in a single proposal distribution.
\paragraph{The same ghost projection.}

All scales still divide higher scales, and
\(1/(c-1)<\delta\), so the guarded child levels are precisely
\(1,\ldots,j-1\).  Allocate the same private-coordinate addresses as
before, now through level \(J+1\).  Ancestor ghosts stop before their
current child's source birth; the youngest stops at the observation
boundary.  The chronology and strictly decreasing levels give the same
absence of distinguished-threshold collisions, including for
off-support proposed descriptors.
Each current threshold is uniform on
\(\{M_i,\ldots,2M_i-1\}\), so the residual channel still has domination
factor three.  Four frame flags give factor twelve.
The positive blind probability ensures that every actual prerequisite
has positive proposal weight.
No other part of the forced-label density calculation changes.
For a live call chain, use the same charge
\[
 \mathcal L_t=
 \log(1/z)+\sum_{\text{frames}}\log\mathcal N
       +\sum_{\text{nonroot frames}}\log(1/p_{i-1}(v))
       +r_t\log12.
\]
In padding use \(\mathcal L_t=\log(1/z_I(e))\) and \(r_t=1\).
There is no residual or flag channel in padding: its mode and complete
state are already specified by the descriptor.  Hence no \(\log12\)
term is needed in that mode.
Both charges are zero in idle and dead states.
\paragraph{Revised causal charges.}

Put
\[
 c_0=1+\log12,\qquad
 \psi_i(v)=\log(1/p_{i-1}(v)),\qquad
 \lambda_i(v)=\log(1+\zeta a_{i-2}(v)).
\]
A nonroot call entry pays
\[
 2K_i[c_0+\psi_i(v)+\lambda_i(v)].
\]
At a root endpoint entry replace \(\psi_i(v)\) by
\(\log(1/z)\), the reciprocal joint prerequisite for that endpoint.
Padding entry pays
\[
 3H'_I[1+\log(1/z_I(e))].
\]
There is no separate root-birth charge.
The first call's entry and its monitor gate are processed at the birth
boundary before that boundary is observed.  A birth followed by an
entry rejection leaves only the dead state.  Thus no root-only state
requires an additional charge.
Every call frame occupies fewer than \(2K_i\) observation boundaries.
Its local-normalizer cost is bounded by \(\lambda_i(v)\), by
\eqref{sharp:eq:child-normalizer}; its anchor unit and residual/flag
cost sum to \(c_0\).  Its prerequisite is \(\psi_i(v)\), except
at the root where it is exactly \(\log(1/z)\).
The call charges therefore cover all integrated call charges.
Padding lasts at most \(3H'_I\) observation boundaries, so its charge
covers its prerequisite logarithm and single anchor.
Charges are accepted before the corresponding state is created, and
cancellation or the external horizon only shortens the charged interval.
Consequently
\begin{equation}\label{sharp:eq:pathwise-coverage}
 \sum_{t<T}(\mathcal L_t+r_t)\le F_{\mathrm{acc}}
\end{equation}
pathwise, with these revised charges.

\begin{proposition}[Sharpened fixed-phase transfer]
\label{sharp:prop:transfer}
Under the standing sharpened parameters, let $G$ be simple with
$m>0$, $1\le T\le m/512$ an integer, $L\ge1$ dyadic, and
$h'\ge c+1$ an integer cap. Let \(V_t\) be a finite auxiliary counter whose updates use only the
step position, pre- and post-step retained core (including a root
completion), its own state, and fresh independent coins. It starts
deterministically, resets on death, and is not consulted by the
sampler. Entropies below are under the random-permutation law.
For the sharpened monitored sampler,
\begin{equation}\label{sharp:eq:transfer}
 \KL(P_T\Vert R_T)
 \le
 \frac{
 T\left[\log(2+\zeta_0\mu'/L)
       +\max_{t<T}\ent(V_t\mid C_t)\right]
       +\E_{\WOR}[F_{\mathrm{acc}}]}{m-T},
 \qquad \zeta_0<12.03.
\end{equation}
\end{proposition}
\begin{proof}
The descriptor likelihood is exactly that of
Lemma~\ref{tr:lem:likelihood}, with the new proposal normalizers.
The ghost construction gives the stated live charges and reconstructs
the complete control state.  The monitor's charges are determined by the
current type and fixed graph parameters and use fresh gates, so the
permutation and i.i.d. states share one causal transition kernel.
Apply Theorem~\ref{tr:thm:anchors}, use
\eqref{sharp:eq:root-normalizer} and
\(\log(m/(m-r_t))\le r_t/(m-t)\), and sum
\eqref{sharp:eq:pathwise-coverage}.
\end{proof}
The expected-charge calculation for the sharpened sampler applies to
exactly these charges.  Combining its bound \(K_0=300\) with
Proposition~\ref{sharp:prop:transfer} gives the comparison used in the
refined phase analysis.  The proof applies directly to the fixed positive choice
\(\theta=2^{-200}\).
\subsection{Sharper phases and the final schedule}
\label{app:sharp-phases}

We now specify the remaining parameter changes that give approximation
factor $2^{28}$. We retain the core and boundary conventions of the
preceding subsection, with fixed blind-pass probability $\theta=2^{-200}$.
In particular, a root born at boundary $b$, with truncated reward $H'$,
finishes at boundary $b+3H'$. Its birth edge is included in its duration.
Successive i.i.d. cycles therefore have lengths $G+3H'$, where $G$ is
geometric on $\{1,2,\ldots\}$ with mean $L$. The reward law is independent
of $G$ and of the phase at which the root is born.
We use the bounds already proved for this core:
\begin{equation}\label{eq:sharp-core-estimates}
 \ell_o\frac{h'}{m'}\nu'\le\mu':=\E[H']\le U\nu',\qquad
 U=635,\qquad \E[F_{\rm root}]\le K_0\mu',\quad K_0=300.
\end{equation}
Here $m'$ and $\nu'$ describe the residual graph, and $h'$ is the
integer reward cap. The current-state information bound has reference
normalizer $2+12.03\mu'/L$. All counters used by the algorithm below
are exact. The monitor and bounded tester are used only in the proof.
\paragraph{One phase.}
Fix its horizon $T$, a dyadic $L\ge1$, and parameters $\kappa,a,b$.
Set $h'=\floor{\kappa T/3}$ and use birth probability $1/L$.
At the final boundary, count a padded root whose completion boundary
is exactly $T$, and start no new root there. Let $N_T$ be this count.
Define
\begin{equation}\label{eq:sharp-output}
 B=2^{24},\qquad \eta=\frac1{256},\qquad
 w=\frac{257}{255}(1+2^{-20})<1.01,\qquad
 Z=\max\left\{1,\frac{(1-a)T}{w(1+3U)(N_T+1)}\right\}.
\end{equation}
The implementation never consults $N_T$ before the final boundary.
Thus this exact counter is excluded from the state comparison, just
as in the simpler analysis.
\paragraph{Renewal concentration.}

Write $y=L+3\mu'$ and $g=T/y$, and suppose $L\le\kappa T/64$.
For $s=u/(\kappa T)$, the geometric moment-generating functions give
\[
 \log\E[e^{sG}]\le\frac{L}{\kappa T}\frac{u}{1-u/64}
 \quad(0<u<64),\qquad
 \log\E[e^{-sG}]\le-\frac{L}{\kappa T}\frac{u}{1+u/64}.
\]
For example, the first MGF is $[1-L(1-e^{-s})]^{-1}$; use
$L(1-e^{-s})\le Ls\le u/64$ and
$-\log(1-z)\le z/(1-z)$. The second follows from
$\log(1+z)\ge z/(1+z)$. Since $0\le3H'\le\kappa T$,
convexity gives
$\log\E[e^{\pm3sH'}]\le(3\mu'/\kappa T)(e^{\pm u}-1)$.
Put
\[
 \psi_+(u)=\max\{e^u-1,u/(1-u/64)\},\qquad
 \psi_-(u)=\min\{1-e^{-u},u/(1+u/64)\}.
\]
For any $u_-\in(0,64)$ and $u_+>0$, Chernoff's inequality yields
\begin{align}
 \Prb_{\IID}\{N_T+1<(1-a)g\}
 &\le p_-:=\exp\left[-\frac{u_--(1-a)\psi_+(u_-)}{\kappa}\right],
 \label{eq:sharp-renewal-lower}\\
 \Prb_{\IID}\{N_T+1>(1+b)g\}
 &\le p_+:=\exp\left[-\frac{(1+b-\kappa-\kappa/64)\psi_-(u_+)-u_+}{\kappa}\right].
 \label{eq:sharp-renewal-upper}
\end{align}
The integer endpoints are useful here. The lower event requires
$\sum_{i=1}^n(G_i+3H'_i)>T$ for $n=\ceil{(1-a)g}-1$, with
$ny\le(1-a)T$. The upper event requires the reverse inequality for
$n=\floor{(1+b)g}$, with
$ny\ge(1+b)T-y\ge(1+b-\kappa-\kappa/64)T$.
If the first value of $n$ is zero, its event is empty.
\paragraph{Testing the count with bounded state.}

An independent counter increments at each completion with probability
$\min\{1,B/g\}$, is capped at $M=\ceil{(1+b)wB}+3$, and is erased
at monitor death. When $g\le B$, accept if its value $V$ satisfies
$(1-a)g\le V+1\le(1+b)g$. When $g>B$, accept if
\[
 (1-\eta)((1-a)B-1)\le V\le(1+\eta)(1+b)B.
\]
The cap lies strictly above the acceptance interval in both cases.
While the monitor is alive, conditional on the sampler and monitor
history, the uncapped counter has law $\operatorname{Bin}(N_T,B/g)$
in the second case. Neither process reads its coins or state.
For $a\le4/5$ and $b\ge1$, Chernoff bounds give the following two
uniform conclusions. A count with
$(1-a)g\le N_T+1\le(1+b)g$ is rejected with probability at most
$\zeta_c<10^{-10}$; a count outside
$[(1-a)g/w,(1+b)wg]$ is accepted with probability at most
$\zeta_s<10^{-10}$. Explicit bounds are
\begin{align*}
 \zeta_c&\le e^{-\eta^2(B/5-1)/2}
                  +e^{-\eta^2B/(2+\eta)},\\
 \zeta_s&\le e^{-\eta^2B/(5w(2+\eta))}
                  +e^{-\eta^2(2wB-1)/2}.
\end{align*}
To check the endpoints in the soundness claim, put $\Delta=2^{-20}$.
The lower offending mean is below $(1-a)B/w$, and
$(1-a)B\Delta\ge(B/5)\Delta\ge1+\Delta$ makes the lower acceptance
threshold at least $(1+\eta)$ times this mean. For the upper offending
mean, $(1+\eta)(1+b)B\Delta\ge1-\eta$ makes the upper threshold
at most $(1-\eta)((1+b)wB-1)$. The tester contributes at most
$\log(M+1)$ to conditional state entropy.
\paragraph{A self-bounding information estimate.}

For $u\in[0,1]$ and $v\in(0,1)$, write
\[
 \operatorname{kl}(u\Vert v)=u\log\frac uv
       +(1-u)\log\frac{1-u}{1-v}
\]
for binary relative entropy, with zero-mass terms interpreted by
continuity. The next argument improves the generic monitor bound by
using its death probability itself to bound accepted charges.

Accept a causal charge $\phi$ with probability $e^{-\phi/(\beta T)}$,
and erase the monitored state on rejection. The accepted charges satisfy
\begin{equation}\label{eq:sharp-monitor-charge}
 \E[F_{\rm acc}]\le\beta T\Prb\{\Dead\}.
\end{equation}
Indeed, $\phi e^{-\phi/(\beta T)}\le\beta T(1-e^{-\phi/(\beta T)})$;
after multiplication by the probability of surviving to this charge,
the rejection terms sum to the probability of first death.
Under i.i.d. input, predictable births and~\eqref{eq:sharp-core-estimates}
bound expected full charges by $K_0(T/3+h')$: completed root rewards sum
to at most $T/3$, and at most one root is unfinished. Thus i.i.d. death
probability is at most $d_0$ when
\[
 \beta=\frac{K_0(1+\kappa)}{3d_0}.
\]
If $\nu'\le LF$, the state information bound and
\eqref{eq:sharp-monitor-charge} imply
\begin{equation}\label{eq:sharp-information}
 X:=\KL(P_T\Vert Q_T)\le t\bigl(c_1(F)+\beta p\bigr),\qquad
 t=\frac{T}{m'-T},\quad p=\Prb_{\WOR}\{\Dead\},
\end{equation}
where $P_T,Q_T$ are the permutation and i.i.d. laws of the monitored
core and tester, and
\[
 c_1(F)=\log(2+12.03UF)+\log\bigl(\ceil{(1+b)wB}+4\bigr).
\]
The following lemma resolves the apparent dependence on the unknown
permutation death probability $p$. A large value of $p$ would require
more binary relative entropy than the affine upper bound allows.

\begin{lemma}[Self-bounding relative entropy]\label{sharp:lem:self-bound}
Let $P,Q$ be probability laws on the same measurable space, and let
$D$ be an event. Write $p=P(D)$ and $X=\KL(P\Vert Q)$.
Suppose $0<d_0<\bar q<1$, $Q(D)\le d_0$, and, for real
$c_1,t,x\ge0$ and $\beta>0$,
$X\le t(c_1+\beta p)$. If
\begin{equation}\label{eq:sharp-information-budget}
 \operatorname{kl}(\bar q\Vert d_0)>x,
 \qquad t(c_1+\beta\bar q)\le x,
\end{equation}
then $p<\bar q$ and $X\le x$.
\end{lemma}
\begin{proof}
Suppose $p\ge\bar q$. Binary data processing and $Q(D)\le d_0<p$
give $X\ge\operatorname{kl}(p\Vert d_0)$. Convexity of
$\operatorname{kl}(\cdot\Vert d_0)$, which vanishes at $d_0$,
then gives
\[
 X\ge\operatorname{kl}(p\Vert d_0)
 \ge\frac{p-d_0}{\bar q-d_0}\operatorname{kl}(\bar q\Vert d_0)
 >x\frac{p-d_0}{\bar q-d_0}
 \ge t(c_1+\beta p).
\]
For the last inequality, multiply the second condition in
\eqref{eq:sharp-information-budget} by
$(p-d_0)/(\bar q-d_0)$ and use
$(p-\bar q)(c_1+\beta d_0)\ge0$. This contradicts the assumed
upper bound on $X$. Hence $p<\bar q$, and substitution into that
upper bound yields $X\le x$.
\end{proof}

Apply the lemma with the laws in~\eqref{eq:sharp-information},
the death event $D$, and $c_1=c_1(F)$. Under the two inequalities
in~\eqref{eq:sharp-information-budget}, it gives $p<\bar q$ and
$X\le x$.
Under i.i.d. input, death or tester rejection has probability at most
$p_0=d_0+p_-+p_++\zeta_c$. If
$\operatorname{kl}(f_0\Vert p_0)>x$ and $f_0>p_0$, binary data
processing bounds its permutation probability by $f_0$.
Tester soundness therefore implies, with failure at most
$f_0+\zeta_s$,
\begin{equation}\label{eq:sharp-count-good}
 \frac{1-a}{w}\frac{T}{L+3\mu'}
 \le N_T+1\le(1+b)w\frac{T}{L+3\mu'}.
\end{equation}
If $L\le\nu_0$, $\nu'\le\nu_0$, and this event holds, then
\begin{equation}\label{eq:sharp-candidate-bound}
 \frac{\nu'T}{Gm'}\left(1-\frac3{\kappa T}\right)
 \le Z\le\nu_0,
 \qquad
 G=\frac{3(1+b)w^2(U+1/3)}{(1-a)\ell_o\kappa}.
\end{equation}
The upper bound uses $L+3\mu'\le(1+3U)\nu_0$.
For the lower bound, discard $L$, use the lower bound on $\mu'$,
and substitute $h'\ge\kappa T/3-1$.
\paragraph{Parameters and arithmetic certificate.}

The two types of phases use the following rational parameters and
strict upper bounds. Decimal entries denote exact terminating decimals.
\begin{center}
\small
\begin{tabular}{@{}lcc@{}}
\toprule
 & Coarse phase & Final phase\\
\midrule
$(\kappa,a,b)$ & $(1/32,1/2,1)$ & $(3/32,5/8,5/4)$\\
$(d_0,x,\bar q)$ & $(1/100,101/12800,1/40)$ & $(3/40,3/10,7/20)$\\
$(u_-,u_+)$ & $(693/1000,677/1000)$ & $(981/1000,768/1000)$\\
$p_-$ & $<0.00207$ & $<0.0225$\\
$p_+$ & $<0.000091$ & $<0.0161$\\
$p_0$ & $<0.0122$ & $<0.114$\\
$f_0$ & $0.03$ & $0.42$\\
$\operatorname{kl}(\bar q\Vert d_0)$ & $>0.00802$ & $>0.3098$\\
$\operatorname{kl}(f_0\Vert\text{displayed }p_0)$ & $>0.00935$ & $>0.3019$\\
\bottomrule
\end{tabular}
\end{center}
Thus coarse-copy failure is below $1/10$, and final-copy failure is
below $17/40$, whenever~\eqref{eq:sharp-information-budget} holds.
For $F\ge1$, $\Lambda(F)=1+\ceil{\log_2 F}$, and $B_1=2^{36}$,
the remaining information estimates are
\begin{equation}\label{eq:sharp-information-constants}
 c_{1,f}(B_1)<52,\qquad
 c_{1,c}(F)+\beta_c\bar q_c<284+\Lambda(F),\qquad
 \frac{3/10}{52+\beta_f\bar q_f}>0.000533.
\end{equation}
Every inequality in this paragraph, including $\zeta_c,\zeta_s<10^{-10}$,
can be checked using rational bounds as follows. For $0\le u\le1$,
bound $e^u$ between its degree-$90$ Taylor polynomial and that
polynomial plus
$u^{91}/(91!(1-u/92))$. To bound $\log x$, write $x=2^jv$ with
$1\le v\le2$, put $z=(v-1)/(v+1)$, and use
\[
 0\le\log v-2\sum_{i=0}^{89}\frac{z^{2i+1}}{2i+1}
 \le\frac{2z^{181}}{181(1-z^2)}.
\]
The same expression at $v=2$ bounds $\log2$. These enclosures give
the displayed strict inequalities using only rational arithmetic.
For the coarse information estimate, it suffices to check
$\log(2+12.03U)+\log(M_c+1)+\beta_c\bar q_c<285$ and use
$\log F\le\Lambda(F)-1$.
\paragraph{Prescribed-length algorithm.}

The following replaces the numerical schedule in the main algorithm;
the phase subroutine is~\eqref{eq:sharp-output}. Return zero if $m_0=0$
and one if $1\le m_0\le C_0:=2^{27}$. Otherwise set $L=1$ and define
\begin{equation}\label{eq:sharp-targets}
 B_1=2^{36},\qquad
 e_s=\floor{\frac{B_{s-1}}{2^{26}2^{s-2}}}-321,\quad
 B_s=2^{e_s}\quad(s\ge2).
\end{equation}
Let $k$ be the least index with $B_k\ge m_0$, and put $F_k=m_0$
and $F_s=B_s$ for $s<k$.
\begin{enumerate}
\item For $s=k,k-1,\ldots,2$, set
\[
 T_s=\floor{\frac{2^{-7}2^{-(s-2)}m_0}{320+\Lambda(F_s)}},
 \qquad r_s=2s+11.
\]
If $2^{11}L\ge T_s$, skip this stage. Otherwise keep this $L$ fixed,
run $r_s$ consecutive coarse phases of horizon $T_s$, and set
$L\leftarrow\max\{L,\dyad(\operatorname{median}(Z_1,\ldots,Z_{r_s}))\}$.
Here $\dyad(z)$ is the largest power of two at most $z$.
\item Set $T_1=\floor{m_0/2048}$. If $L\ge3T_1/2048$, return $L$.
Otherwise keep $L$ fixed, run nine consecutive final phases of horizon
$T_1$, and return $\max\{L,\operatorname{median}(Z_1,\ldots,Z_9)\}$.
\end{enumerate}
These instructions involve only integers and rational comparisons.
There is no need to form a target exceeding $m_0$: compare its exponent
with $\ceil{\log_2m_0}$ instead. For each coarse stage one may store
the exponents of $\dyad(Z_i)$, since dyadic rounding commutes with an
odd median. All phase horizons, caps, birth rates, and stopping rules
are therefore computable within the stated memory bound.
\paragraph{Deterministic bounds for the schedule.}

First $e_2=703$, and $e_s\ge2e_{s-1}$ for $s\ge3$.
For an explicit induction, $e=e_{s-1}\ge703$ and $e\ge2s+48$ imply
$e_s=2^{e-s-24}-321\ge2^{\floor{e/2}}-321\ge2e$.
The last inequality holds at $e=703,704$ and propagates in steps of two;
the induction also preserves $e_s\ge2(s+1)+48$.
Hence $k=O(\log\log(m_0+2))$.
If a coarse stage $s$ occurs, $q=\log_2m_0>36$ and
$q\ge36\cdot2^{s-2}$. Consequently
\[
 T_s\ge\frac{36m_0}{128(322+q)q}-1>10^6.
\]
For the last inequality, $2^q/[q(q+322)]$ increases for $q\ge36$,
and the bound at $q=36$ exceeds $10^6$.
Thus every coarse cap is at least $513$. Since $m_0>C_0$, final
horizons satisfy $T_1\ge65536$, and final caps are at least $2048$.
The skip rules ensure $L<\kappa T/64$ whenever a phase is run.
All phases finish within the first $\epsilon m_0$ positions, where
\begin{equation}\label{eq:sharp-prefix}
 \sum_{s\ge2}\frac{(2s+11)2^{-(s-2)}}{128(320+37)}
 +\frac9{2048}
 =\frac{221}{43008}<\epsilon:=\frac{11}{2048}.
\end{equation}
In particular, $m'-T\ge(1-\epsilon)m_0$ for every phase.
For coarse phases, $(1-\epsilon)^{-1}<1.01$ and
\eqref{eq:sharp-information-constants} give
\[
 \frac{T}{m'-T}(c_1(F_s)+\beta_c\bar q_c)
 <\frac{1.01}{128}\,
   \frac{2^{-(s-2)}(284+\Lambda(F_s))}{320+\Lambda(F_s)}
 \le\frac{101}{12800}.
\]
For final phases,
$T/(m'-T)\le1/[2048(1-\epsilon)]<0.000491$, which is below
the permitted ratio in~\eqref{eq:sharp-information-constants}.
These bounds also imply $T/m'<1/512$, as required by the sharper
reference comparison.
The cap-to-size ratios are also at most $1/49152$.
Substituting this bound and the core constant
$\kappa_{\rm low}\le11/10$ into the lower proxy estimate, with its
load parameter set to $1/256$, gives
\[
 \ell_o\ge
 \frac{(255/256)^2}
 {1+2(255/256)(11/10)(1/49152)/(1/256)}
 =\frac{65025}{66284}>\frac{49}{50}.
\]
Thus the coefficients in~\eqref{eq:sharp-candidate-bound} satisfy
$G_c<254000$ and $G_f<127000$.
\paragraph{Correctness with prescribed length.}

Let $\nu_0$ denote the matching size of the prescribed $m_0$-edge graph.
Maintain $1\le L\le\nu_0\le LF_s$ at the start of stage $s$.
Suppose every residual graph has matching size at least
$(127/128)\nu_0$. If a strict majority of a coarse stage's phases
satisfy~\eqref{eq:sharp-count-good}, then
\[
 \frac{\nu_0}{L_{\rm new}}
 \le2^{26}2^{s-2}(320+\Lambda(F_s))
 \le2^{26}2^{s-2}(321+e_s)\le B_{s-1}.
\]
Indeed dyadic rounding costs at most two, and the residual-matching,
horizon-floor, and cap-floor losses together are bounded by
\[
 \frac{128}{127}\frac{1+10^{-6}}{1-96\cdot10^{-6}}<1.02;
 \qquad 2\cdot254000\cdot1.02\cdot128<2^{26}.
\]
A skipped stage has a smaller ratio, using $L\ge T_s/2048$.
The upper invariant follows from the upper bound on each good candidate
and does not require residual matching preservation.
A good final majority gives a factor at most
\begin{equation}\label{eq:sharp-known-factor}
 C_{\rm kn}=127000\cdot2048\cdot\frac{128}{127}\cdot\frac{1002}{1000}
 =262668288.
\end{equation}
Here the horizon and cap floors cost less than
$(32768/32767)(1024/1023)<1002/1000$.
A skipped final stage gives the smaller factor
$(2048^2/3)(1+1/65536)$, and the small-input branch is covered because
$C_0<C_{\rm kn}$.
The phase estimates apply conditionally on each phase's complete past.
To avoid conditioning on later matching preservation, define an
artificial copy-failure indicator to be zero if its stage invariant
fails, and otherwise to indicate failure of~\eqref{eq:sharp-count-good}.
These indicators are stochastically dominated, conditionally on the
past, by Bernoulli variables of parameter $1/10$ in coarse stages
and $17/40$ in the final stage. Therefore the total probability of a
bad coarse majority is at most
\[
 \sum_{s\ge2}(3/5)^{2s+11}
 =\frac{14348907}{19531250000}<0.00075,
\]
and final-majority failure is at most
\[
 \sum_{j=5}^9\binom9j(17/40)^j(23/40)^{9-j}
 =\frac{8411044027019}{26214400000000}<0.321.
\]
For $\nu_0>C_0$, the number of edges of a fixed maximum matching
among the first $\floor{\epsilon m_0}$ positions is hypergeometric,
with mean and variance at most $\epsilon\nu_0$. Since
$1/128-\epsilon>1/512$, matching preservation fails with probability
at most $\epsilon2^{18}/C_0=11/2^{20}$.
For $\nu_0\le C_0$, the fallback value one already gives the lower
bound; good phases preserve the upper bound without this event, since
$\nu_0\le C_0<F_s$ for every executed stage and $L\ge1$.
\paragraph{Unknown stream length.}

Use the same wrapper as in the main proof, with
$\xi=\eta_u=3/512$. Set $s_1=1$ and
$s_{j+1}=s_j+\ceil{\xi s_j}$. At global boundary $s_j$, start a
fresh prescribed-size-$s_j$ instance; commit its answer at boundary
$s_{j+1}$, before starting the next instance. The intervals are
disjoint, and~\eqref{eq:sharp-prefix} puts every instance's reads
within its interval. Return the last committed answer, with fallback
one on a nonempty stream and zero on an empty stream.
For analysis, let $s$ be the last committed size; it is deterministic
given $m$. The whole committed interval has deterministic length
$\ceil{\eta_u s}\le s$. Extend the edges in that interval, in their
order, by uniformly sampling an ordered completion from the other
edges until $s$ edges have been chosen. The resulting $H$ is a uniform
$s$-edge subgraph, with uniform order conditional on $H$. Its initial
interval contains every edge read by the prescribed-length algorithm,
so that algorithm has exactly the committed answer. This construction
does not condition on its random stopping time and requires no union
bound over earlier instances.
Write $\nu=\nu(G)$. If $\nu>C_0$, the next commitment has not occurred,
and hence
\[
 m<(1+\xi)^2s+2+\xi,\qquad s>C_0/2,\qquad
 p:=\frac{s}{m}>
 \frac1{(1+3/512)^2+5/C_0}>\frac{505}{512}.
\]
For a fixed maximum matching of $G$, its number $X$ of edges in $H$
has mean $p\nu$ and variance below $(7/512)\nu$.
Since $p-63/64>1/512$,
$\Prb\{X<(63/64)\nu\}\le7/2^{18}$ by Chebyshev's inequality.
Combined with the prescribed-length preservation event, this costs
at most $39/2^{20}<0.00004$. If $\nu\le C_0$, the fallback one
again suffices for the lower bound, and the upper-bound argument on
$H$ applies without either preservation event. Thus the success
probability is greater than
\[
 1-0.321-0.00075-0.00004>2/3,
\]
and the approximation factor is at most
\[
 C_{\rm kn}\frac{64}{63}
 =\frac{5603590144}{21}<2^{28}.
\]
Round the final rational answer upward to an integer. On the good
event it is at most the integer $\nu(G)$, so rounding preserves the
upper bound and improves the lower bound.
Only one recursive stack and one instance are active. The recursive
core still uses $O(\log^2(n+2))$ bits on every execution. Target exponents,
the phase counter, and $O(\log\log(m+2))$ candidates fit in this
bound. All boundary operations and rational comparisons have the
same word-RAM implementation as before. Finally
$s_{j+1}\ge(1+\xi)s_j$ and each started size is at most $m$, so
$\sum_j s_j\le(1+\xi)m/\xi$. The wrapper therefore preserves the
total running-time bound $O((m+1)\log(n+2))$.
This proves Theorem~\ref{sharp:thm:main}.

\paragraph{Use of AI tools.}
This work resulted from extensive interaction between the authors and OpenAI Astra. Throughout an iterative process, the authors provided suggestions, critically evaluated Astra's responses, and refined the proposed ideas and arguments. OpenAI Astra was also used to assist with writing and revising the manuscript based on detailed prompts provided by the authors, with Gemini Argon contributing to selected parts of the process. The authors directed the entire process and critically assessed all proposed ideas and arguments. Lean files formalizing the main theorem are available at https://github.com/hajiaghayi/Random-Matching.

\bibliographystyle{plain}
\bibliography{random_order_matching_final}
\end{document}